\documentclass[aps,twocolumn,superscriptaddress,groupedaddress,amsmath,amssymb]{revtex4-2}
\usepackage{graphicx}
\usepackage{dcolumn}
\usepackage{bm}        
\usepackage{amssymb}   

\usepackage{bm}
\usepackage[usenames ,dvipsnames]{xcolor}
\usepackage{color}
\usepackage{url}
\usepackage{dsfont}        
\usepackage{slashed}
\usepackage{epsfig}
\usepackage{multirow,array}
\usepackage{float}
\usepackage{fancyhdr}
\usepackage{indentfirst}
\usepackage{cancel}
\usepackage{tabularx}
\usepackage{simplewick}
\usepackage{amsmath}
\usepackage[normalem]{ulem}
\usepackage{slashed}
\usepackage{upgreek}
\usepackage{xtab,afterpage,longtable}
\RequirePackage[colorlinks=true
  ,urlcolor=Blue
  ,anchorcolor=blue
  ,citecolor=orange
  ,filecolor=blue
  ,linkcolor=blue
  ,menucolor=blue
  ,pagecolor=blue
  ,linktocpage=true
  ,pdfproducer=medialab
  ,pdfa=true
]{hyperref}

\def\beq{\begin{equation}}
\def\eeq{\end{equation}}
\def\bea{\begin{eqnarray}}
\def\eea{\end{eqnarray}}
\begin{document}

\title{Dynamics of Long-lived Axion Domain Walls and Its Cosmological Implications}
     
\author{Chia-Feng Chang}
\email{chiafeng.chang@email.ucr.edu}
\author{Yanou Cui}
\email{yanou.cui@ucr.edu}
\affiliation{Department of Physics and Astronomy, University of California, Riverside, CA 92521, USA}

\date{\today}

\begin{abstract}
We perform an updated analysis on a long-lived axion domain wall (DW) network. By simulating the axion field on a 3D lattice and fitting an analytical model for the DW evolution, we identify the leading energy loss mechanisms of the DWs and compute the spectrum of axions emitted from the network. The contribution from the DWs to axion dark matter (DM) density is derived, with viable parameter space given. The application to both QCD axions and general axion-like particles (ALPs) are considered. Due to the new approaches taken, while our results bear consistency with earlier literature, notable discrepancies are also revealed, such as the prediction for DM abundance, which may have a profound impact on axion phenomenology at large.
\end{abstract}

\maketitle

\section{Introduction}
Axions are ultra-light particles that are originally proposed as a compelling solution to the Strong CP problem in quantum chromodynamics (QCD) \cite{Peccei:1977hh,Peccei:1977ur,Wilczek:1977pj}. Recent years have seen a significantly increased interest in QCD axions and more general axion-like particles (ALPs), as dark matter (DM) candidates alternative to WIMPs \cite{Weinberg:1977ma,Abbott:1982af,Dine:1982ah,Preskill:1982cy}. While most existing studies on axion phenomenology and detection focused on the axion particle per se, the impact of the accompanying axion topological defects, i.e.~axion strings and domain walls (DWs), can be substantial, yet still not well understood. Such axion topological defects are indispensable companions of axion particles for post-inflationary PQ symmetry breaking, with potentially significant contribution to axion relic abundance \cite{Sikivie:1982qv,Vilenkin:1984ib,Davis:1986xc,Vincent:1996rb,Kawasaki:2014sqa,Vilenkin:1986ku}, and may provide complementary search avenues for axion models \cite{Marsh:2015xka,Borsanyi:2015cka,Hlozek:2014lca,Kawasaki:2013ae,Chang:2019mza,Chang:2021afa,LISACosmologyWorkingGroup:2022jok,Brzeminski:2022haa,Agrawal:2020euj,Jain:2021shf,Jain:2022jrp,Agrawal:2019lkr,Dessert:2021bkv}. A growing effort has been made in the past few years along this direction. However, there are still debates to be resolved and clarifications to be made, in part due to the technical challenges with simulating axion topological defects \cite{Klaer:2017qhr,Kawasaki:2018bzv,Martins:2018dqg,Hindmarsh:2019csc,Hook:2018dlk,Buschmann:2021sdq,Gorghetto:2020qws,Vaquero:2018tib,Gorghetto:2018myk,Buschmann:2019icd,Hiramatsu:2010yu,Hiramatsu:2012sc,Hiramatsu:2012gg}.

Axion cosmic strings form as the PQ breaking phase transition (PT) occurs at a high energy scale $f_a$, and prevail till the pseudo-goldstone boson (axion) later acquires a nonzero mass $m_a$ and DWs form. The structure of the DWs depends on the model specifics of the axion potential and is characterized by the axion mass and the DW number $N_{\rm DW}$. The case with $N_{\rm DW}=1$ is most studied in recent years, where the DWs are short-lived and strings dominate the dynamics of the axion topological defects \cite{Buschmann:2019icd,Buschmann:2021sdq,Vaquero:2018tib}. On the other hand, more generally for the $N_{\rm DW} > 1$ models e.g.~Dine-Fischler-Srednicki-Zhitnitsky model \cite{Zhitnitsky:1980tq,Dine:1981rt}, the DWs are stable and problematic as they would over-close the Universe. Nevertheless, the $N_{\rm DW}>1$ cases can be innocuous with the presence of a small symmetry-breaking bias term in the axion potential, which yields the DWs that are long-lived but collapse before the BBN \cite{Zeldovich:1974uw,Saikawa:2017hiv}. Upon collapsing, long-lived DWs can leave observable imprints in the form of axion dark matter relic density, gravitational waves (GWs), as well as the impact on cosmic structure formation \cite{Kawasaki:2014sqa,Hiramatsu:2013qaa}. A clear understanding of the evolution and dynamics of the DW network is crucial for predicting and probing such potentially rich phenomenology. However, the literature on the dynamics of metastable DWs (axion-associated or more general) is still relatively scarce \cite{Hindmarsh:1996xv,Hindmarsh:2002bq,Martins:2016ois,Correia:2018tty,Correia:2014kqa,Kawasaki:2014sqa,Hiramatsu:2013qaa}, and further investigation is required to advance and clarify our understanding.
 
In this work, we conduct an updated analysis for the long-lived axion DWs and predict axion relic abundance produced from the axion DWs (with $N_{\rm DW}$=2 as a benchmark). We perform a 3D field theory lattice simulation for the axion field with grid size $N^3 = 1536^3$ in a radiation-dominated background, including a bias term in the axion potential, and solve the axion field equation of motion exactly. This differs from earlier simulation work, with the promise of potential improvement: e.g.~the analysis of metastable DWs in \cite{Kawasaki:2014sqa} and \cite{Hiramatsu:2012sc} is based on a 2D simulation, while the 3D simulation in \cite{Martins:2016ois,Correia:2018tty} employs Higgs DWs with Press-Ryden-Spergel (PRS) \cite{Press:1989yh} approximation. In order to elucidate the physics of the dynamics of DW evolution, we investigated the DW radiation mechanisms by capturing and zooming in the snapshots of animations from our simulation and by analyzing the axion spectrum and zoom-in. In addition to obtaining results based on numerical simulation, through analytical fitting, we also present the velocity-dependent one-scale (VOS) model applicable to the metastable DW evolution. This is a notable extension of the framework of the VOS model which previously has been widely used to describe the evolution of other types of topological defects such as cosmic strings \cite{Martins:1996jp,Martins:2000cs} and, only recently a few attempts on stable DWs \cite{Martins:2016ois,Martins:2016lzc, Correia:2018tty,Avelino:2005kn,Leite:2011sc,Leite:2012vn}. By combining numerical and analytical approaches, our analysis leads to an updated prediction for the spectrum and relic abundance of axions radiated from DWs, as well as new insights into the evolution of DW substructures. This study may shed new light on the cosmological implication of axion topological defects and their role in axion physics at large. 
 
 In the following, we will first introduce the axion model and simulation setup that we adopted. Then we will present the essential results on the dynamics of axion DWs derived from the simulation, and demonstrate how these can be used to calibrate the analytical VOS model. Cosmological implications related to axion DM will be discussed before we conclude.

\section{\label{Sec: Axion Model}Axion model}

We first introduce the benchmark axion model that we consider and the essentials in our simulation. As a pseudo-Nambu-Goldstone boson, axion is associated with the angular mode of a complex scalar field whose VEV spontaneously breaks a global U(1) symmetry. The U(1) symmetry breaking occurs at a relatively high scale $T\sim f_a$ when the radial mode acquires a mass $m_R \sim f_a$. The original shift symmetry possessed by the axion is broken at a much later time $T\sim\Lambda\simeq \sqrt{m_a f_a}$ (e.g.~$\Lambda_{\rm QCD}$ for QCD axion), when the axion acquires a mass $m_a$ and DW forms. At an even later time when $H\ll f_a$, the effective Lagrangian for axion field $a=a(\textbf{x},t)$ with the radial mode integrated out reads
\begin{align}
    \mathcal{L} = |\partial_\mu a|^2 - V(a).
\end{align}
We consider a biased potential
\begin{align}
\label{Eq: Potential}
        V(a) = \frac{m_a^2 f_a^2}{N_{\rm DW}^2} \Bigg[ 1 - \cos\left(N_{\rm DW} \frac{a}{f_a}\right) + \epsilon \left( 1 + \cos\frac{a}{f_a}\right) \Bigg],
\end{align}
where $\epsilon \ll 1$ is the bias parameter that causes the DW to collapse. We consider $N_{\rm DW} = 2$, which implies one true vacuum and one false vacuum in the model \footnote{It is worth mentioning that the bias term in Eq.(\ref{Eq: Potential}) doesn't shift the true vacuum in the axion potential, which is for avoiding the axion quality problem, see a review in \cite{Hook:2018dlk}.}.
This is a representative choice that involves a simple DW structure which eases the simulation analysis and also allows us to extrapolate our results to the string-wall scenario, which we will discuss in more detail in the Appendix A. 

 We estimate the DW surface tension based on the axion potential in Eq.(\ref{Eq: Potential}) as: 
\begin{align}
    \sigma_{\rm DW} \simeq \eta_{\rm DW} \, \frac{m_a f_a^2}{N_{\rm DW}^2},
\end{align}
where $\eta_{\rm DW} = 8$ for the potential in Eq.(\ref{Eq: Potential}), which we will use in this study. $\eta_{\rm DW} = 8.97(5)$ for QCD axion with pion contribution included \cite{GrillidiCortona:2015jxo}. The DWs become dynamical at cosmic time $t\sim 1/m_a$ when the horizon becomes comparable to the DW thickness $\delta \sim 1/m_a$.

\section{\label{Sec: Setup} Simulation} 

\subsection{\label{Sec: Simulation Setup}Setup}

The equation of motion (EoM) of the axion field in a flat homogeneous and isotropic Friedmann-Lemaitre-Robertson-Walker (FLRW) universe is
\begin{align}
\label{Eq: Axion EoM}
     \frac{\partial^2 a}{\partial \tau^2} + 2 \left( \frac{d \hbox{ln} R}{d \hbox{ln}\tau} \right) \frac{1}{\tau} \frac{\partial a}{\partial \tau} - \frac{\partial^2 a}{ \partial x_i^2} = - R^2 \frac{\partial V}{\partial a},
\end{align}
where $R(t)$ is the scale factor, $x_i$ is comoving spatial coordinate, $\tau$ is comoving time, and $\nabla$ is the Laplacian in physical coordinates. We start our simulations at a time that is slightly earlier than the DW formation time. 

For the initial condition (IC) of the field of our simulation, a random and uniform distribution of the axion field is consistent with the consequence of stochastic inflation under the assumption that the axion potential scale $\sqrt{m_a f_a}$ is far below the inflation scale $H_I$ (see \cite{Graham:2018jyp} and a review of the stochastic method \cite{Markkanen:2019kpv}). We thus consider a simpler scenario in that we randomly assign field value $a = 0$ or $\pi$ (the two vacuums in the potential) to realize an unbiased IC where half of the points on the lattice are in a true vacuum and assume zero initial field velocity $\dot{a}(t_i) \to 0$. As we will see once the DW network enters the attractive solution, the so-called scaling regime, the DW network evolution would no longer be sensitive to IC. This phenomena has been observed in earlier simulations \cite{Kamionkowski:2014zda,Kawasaki:2018bzv,Kawasaki:2013ae,Avelino:2005kn,Hiramatsu:2013qaa}, and see \cite{Correia:2018tty} for a discussion of the effect of a biased IC on the PRS DW evolution (and earlier references \cite{Correia:2014kqa,Coulson:1995nv,Larsson:1996sp}). 

Other simulation setups are as follows. We normalized all parameters according to $f_a \to 1$. The lattice size is $N^3 = 1536^3$, and the simulation period starts from $1/H(t_i) = R(t_i) \Delta x_i$, and ends at $1/H(t_f) = (N/2) \Delta x_f$, where $\Delta x_i = 1$ is initial lattice spacing, $R(t_i) = 1$ is initial scaling factor, $\Delta x_f = R(t_f)\Delta x_i$ is comoving spacing at the end of simulation, and a radiation background is assumed with $R(t)\propto t^{1/2}$. We fix the time interval $\Delta \tau = 0.1$ and test convergence by re-running with smaller time intervals, where $\tau$ is the comoving time. We further fix the physical DW thickness as 
\begin{align}
\label{Eq: delta}
    \delta  \sim \frac{1}{m_a} = \frac{1}{(N/2)R(t_i)\Delta x_i}.
\end{align}
These choices imply that the simulation starts at the time when the horizon size equals lattice spacing $\Delta x_i$, and ends when the horizon expands to half of the full lattice size. On the other hand, the DW thickness $\delta$ occupies $N/2$ lattice grids at $t_i$, then as the coordinate expands, the simulation ends when $\delta$ occupies two grids. We chose such simulation setups for the following reasons: \\
(1) $\delta$ cannot be smaller than the size of two grids for sufficient resolution of the DW. Lower resolution leads to incorrect and insensible simulation results such as a frozen DW in the lattice because the gradient $\nabla^2 a$ in the equation of motion Eq.(\ref{Eq: Axion EoM}) would be incorrectly calculated in the simulation. In addition, a lower resolution would incorrectly induce a wrong tail in the axion kinetic spectrum around axion momentum of $k\sim 2\pi/\Delta x_f$. \\
(2) We simulated with two types of boundary conditions (b.c.'s), periodic and symmetric, and investigated the results' robustness against the choice of b.c. As the simulation results are expected to be inevitably subject to b.c. (albeit not significantly as we found), in order to mitigate the effect we conservatively collect simulation data from the central $1/8$ of the simulation box and discard the rest. This data collection range equals the Hubble box size at the end of the simulation. 

In order to present a free axion spectrum by filtering out the DW contribution, we employ a mask function on the axion field as in previous studies \cite{Hiramatsu:2010yu,Kawasaki:2014sqa} (originally applied in CMB analysis \cite{WMAP:2003zzr}). The method is to mask $\dot{a}(x)$ by a window function
\begin{align}
\label{Eq: Mask function}
    \dot{a}(x) \to \theta(x-d)\dot{a}(x),
\end{align}
where $x$ is the coordinate that origin at the DW center where $V(a(x=0)) = V_{\rm max}$, $d$ is a mask function parameter, and $\theta(x)$ is the Heaviside step function. We fix $d = \delta/2$ in our simulation to exclude the DWs contribution to the power spectrum. But due to the influence on the DWs exerted by the background axion field, $\delta$ would not be perfectly a constant. Thus we cannot fully erase the DW contribution to the free axion spectrum, yet our approach should provide a good estimation. A more effective algorithm to erase such a contribution may be developed with dedicated future work. The kinetic power spectrum is found to be insensitive to the choice of $d$ that is not too far from $\delta$, i.e.~$\delta / 4 \lesssim d \lesssim 2 \delta$. We found that applying the mask function on the axion field itself $a(x) \to \theta(x-d)a(x)$ causes an insensible result on the gradient energy and potential, i.e. a variation on the blue tail of spectrum ($k \sim 1/m_a$) sensitive to the variation of $d$. This may be caused by the oscillation behavior of the axion field around the vacuum such as the contribution from sub-horizon compact DW or oscillons (see the red points at the end of the simulation, i.e. the far right panel in Fig.~\ref{Fig: Simulation}) that cannot be fully removed by the mask function. Thus to estimate the total energy of the radiated free axions we only apply the mask function for the axion kinetic energy and assume that the free axions are all in harmonic mode i.e.~its kinetic energy takes half of its total energy . 

Our DW simulation was run with various simulation conditions and ALP model benchmarks as follows. We conducted 5 simulations for each benchmark with $\epsilon \gtrsim 10^{-3}$ (to ensure that all the DWs decayed away by the end of simulation) while keeping the aforementioned parameters constant as described in last three paragraphs. Subsequently, based on the simulation data, we will construct a model for the DW dynamics and then extrapolate it to lower $\epsilon$ values and a wider range of $m_a$ via analyzing the axion spectrum as well as monitoring the evolution of the DWs and the free axion background field informed by the snapshots of simulation and the spectrum analysis in Sec.~\ref{Domain wall dynamics}.  

Besides the main simulation runs, we also conducted test runs under various conditions and ALP model benchmarks to ensure that our analysis result would not be affected by the specific simulation parameters that we have set. In particular, the test runs are set as the following. We assessed the impact of varying simulation parameters (with 5 testing runs for each benchmark as well) such as axion mass $m_a$, spanning a range from 0.5 to 2, initial scaling rate $R(t_i)$ with values of 0.5, 1, and 2, and $x_i$ with values of 0.1, 1, and 10. Additionally, we considered different lattice sizes $N$ (512, 1024, and 1536) and the mask function parameter $d$ as previously mentioned. As expected for free axion spectrum as shown in Sec.~\ref{Domain wall dynamics}, and consequently, our conclusions remained unaffected.

\subsection{\label{Sec: Application of our simulation to other models}Application to Other Models}

Although we simulated a network for a simple DW model, our results can be applied to a variety of more complex models if they satisfy the following conditions:\\
(1) The DW network has enough time to enter the scaling regime before its decay. For instance, in our model a large $\epsilon \gtrsim 5 \times 10^{-3}$ (see Sec.~\ref{Sec: Scaling behavior}) would cause the false vacuum to collapse too early for the DW area to have time to converge to a constant, i.e.~enter the scaling regime (see Sec.~\ref{Sec: Scaling behavior} and Eq.(\ref{Eq: A_v}) therein).\\
(2) Essential properties of the DW should be (approximately) the same as in our simulation. For instance, the DW thickness $\delta$ should be kept as a constant during the scaling regime and before the DW starts to decay. Meanwhile, DW number should be $N_{\rm DW} = 2$ as considered in this study. \\

The first condition eliminates the dependence on the DW initial distribution effect when applied to different models. The second ensures that the DW dynamics are congruent with our findings. As an example, in the following, we explain how our simulation can apply to certain QCD axion models. Firstly, a simple condition for a QCD model to be mimicked by our DW-only simulation is for the DW structure to be absent from the model, which can be satisfied in the scenario of a pre-inflationary PQ symmetry breaking or if the vacuum manifold after the PQ phase transition is simply connected (see later discussion in this section and Appendix.~\ref{Appendix-QCD} for a more complex case: a possible application to QCD models with cosmic strings). Secondly, the QCD axion model needs to have the same $N_{\rm DW} = 2$ and the presence of a nonzero $\epsilon$ term in order to avoid the DW over-closure problem. Furthermore, the DW thickness in the QCD model needs to be effectively constant during the simulation time window. Consider that unlike in the model we considered in Sec.~\ref{Sec: Cosmological implication} where $m_a$ and thus DW thickness is a constant, in QCD the DW thickness generally takes a time-dependent form as
\begin{align}
\label{Eq: QCD axion model}
    \frac{1}{\delta_{\rm QCD}} \simeq m_a(T) \simeq \left\{           
\begin{aligned}
m_a \left(\frac{\Lambda_{\rm QCD}}{T} \right)^4 \;\; \hbox{for} \;\; T > \Lambda_{\rm QCD},\\
m_a   \;\;\;\;\;\;\;\; \hbox{for} \;\; T \leq \Lambda_{\rm QCD},
\end{aligned}
\right.
\end{align}
where the QCD scale $\Lambda_{\rm QCD} = 400\,$MeV, $T$ is the cosmic temperature, and the expression is derived from a diluted instanton gas approximation \cite{Burger:2018fvb,Petreczky:2016vrs,Bonati:2018blm,Gorghetto:2018ocs} (also see the results from lattice simulation \cite{Borsanyi:2016ksw,GrillidiCortona:2015jxo}). The QCD axion DW thickness $\delta_{\rm QCD}$ approaches a constant at the transition time $t_a$ when $T\simeq\Lambda_{\rm QCD}$, and afterwards, the QCD axion DW would evolve as in our simulation. \\

We did not simulate a time-dependent thickness $\delta_{\rm QCD}$ due to the computational limitations imposed by the lattice. The DW thickness, which rapidly shrinks as $\delta_{\rm QCD} \propto R(t)^{-4}$ in Eq.(\ref{Eq: QCD axion model}), imposes a significant demand on the evolution time range in our simulation, because the thickness should be at least larger than the lattice spacing for accurate resolution. Due to this limitation, we choose to focus on simulating the cases where $\delta$ can be treated as a constant. In order for our model to approximate a $\delta_{\rm QCD}$ during the simulation time window, we should consider a small $\epsilon \lesssim 10^{-4}$ such that the DW can live long enough to enter the scaling regime after $t_a$. We will discuss the concrete application of this condition on the parameter space in Fig.~\ref{Relic_Plot} in Appendix.~\ref{Appendix}.

In addition to the issue of constant vs. time-dependent DW thickness as discussed above, another key potential difference between our simple model and the QCD case is that some QCD axion models may also involve cosmic strings in the axion topological defect structure, such as in the scenario of post-inflationary $U(1)$ symmetry breaking, where QCD axion strings persist until DW formation. In such a case with $N_{\rm DW}=2$, two DWs attach to a single cosmic string, forming a string-wall network that differs significantly from what the DW-only structure that we considered in our study. Nevertheless, we find that the influence of cosmic strings is negligible when the DW tension dominates the network \cite{Battye:1997jk}, specifically when the condition
\begin{align}
\label{Eq: String-Wall tension}
\sigma_{\rm DW} t/\mu > 1
\end{align}
is satisfied, where $\mu \simeq 2\pi^2 f_a^2 \hbox{ln}(t f_a)$ is the cosmic string tension. Under this condition, the string-wall structure is well approximated by our simulation. However, for higher values of $N_{\rm DW}>2$, where multiple DWs attach to a single string, a more complex scenario arises with the attachment of multi-DWs. We have chosen to leave the investigation of such complex scenarios with $N_{\rm DW}>2$ for future work. We will present the viable parameter space satisfying the condition given in Eq.(\ref{Eq: String-Wall tension}), and discuss the application to the QCD axion model with cosmic strings in the Appendix.\ref{Appendix-QCD}. 

Furthermore, our decision to focus on the simplified case without string contribution is also influenced by technical considerations. Due to limitations in our simulation resources, the lattice size imposes constraints on extending the simulation period sufficiently to observe DW decay if cosmic strings are included. The scale hierarchy between the width of the string ($\sim 1/f_a$) and the Hubble scale at the time of DW decay prevents us from adequately observing the network in our simulation with the current lattice size.

Finally, note that our simulation results not only can apply to the aforementioned QCD axion models, but also to other axion-like particle models that satisfy the two conditions that we identified above.

\section{\label{Domain wall dynamics}Domain wall dynamics}

\begin{figure*}[t]
 \includegraphics[width=1.03\textwidth]{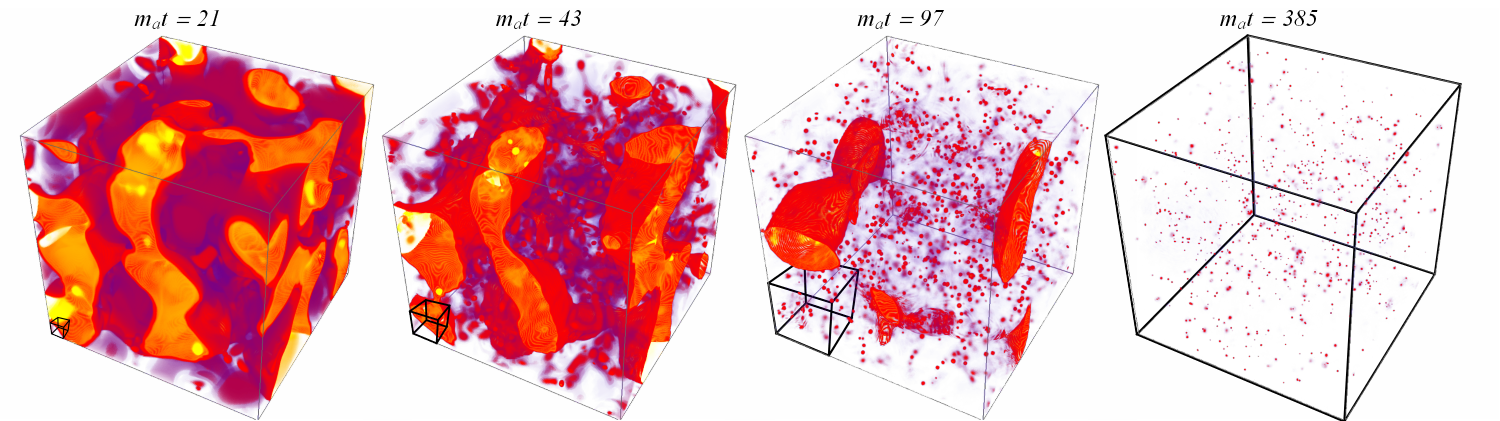}
  \caption{\label{Fig: Simulation} Visualization of lattice simulation with bias parameter $\epsilon = 0.0013$: snapshots in a time series (left to right: $m_a t = 21, 43, 97, 385$). The yellow (blue) region indicates a false (true) vacuum, and the red region represents DWs. The Hubble volume is shown as a black cube in the bottom-left corner of each snapshot (see \href{https://drive.google.com/file/d/1Srfs_sLM4BDl0TCTiAlLsTSCOPCO69Nk/view?usp=sharing}{animation} for $\epsilon = 0.0012$). The small red dots are defined as sub-horizon compact DW or oscillon, which are the axion field that oscillates around the false vacuum, surrounded by DWs (For a zoomed-in simulation for the dissipation of the small red dots, see: \href{https://drive.google.com/file/d/1agukZIEx-G6tT8jW5Ss0iyfJAtobehZs/view?usp=sharing}{animation link})}
\end{figure*}

\begin{figure}[t]
\includegraphics[width=0.465\textwidth]{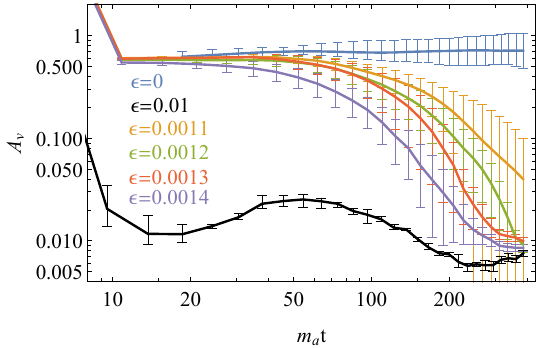} 
\caption{\label{Fig: AVt_mat} DW area parameter (defined in Eq.(\ref{Eq: A_v})) as a function of the cosmic time in our simulation, with varying bias parameter $\epsilon$ (defined in Eq.(\ref{Eq: Potential})).}
\end{figure}

\begin{figure*}[t]
 \includegraphics[width=1.03\textwidth]{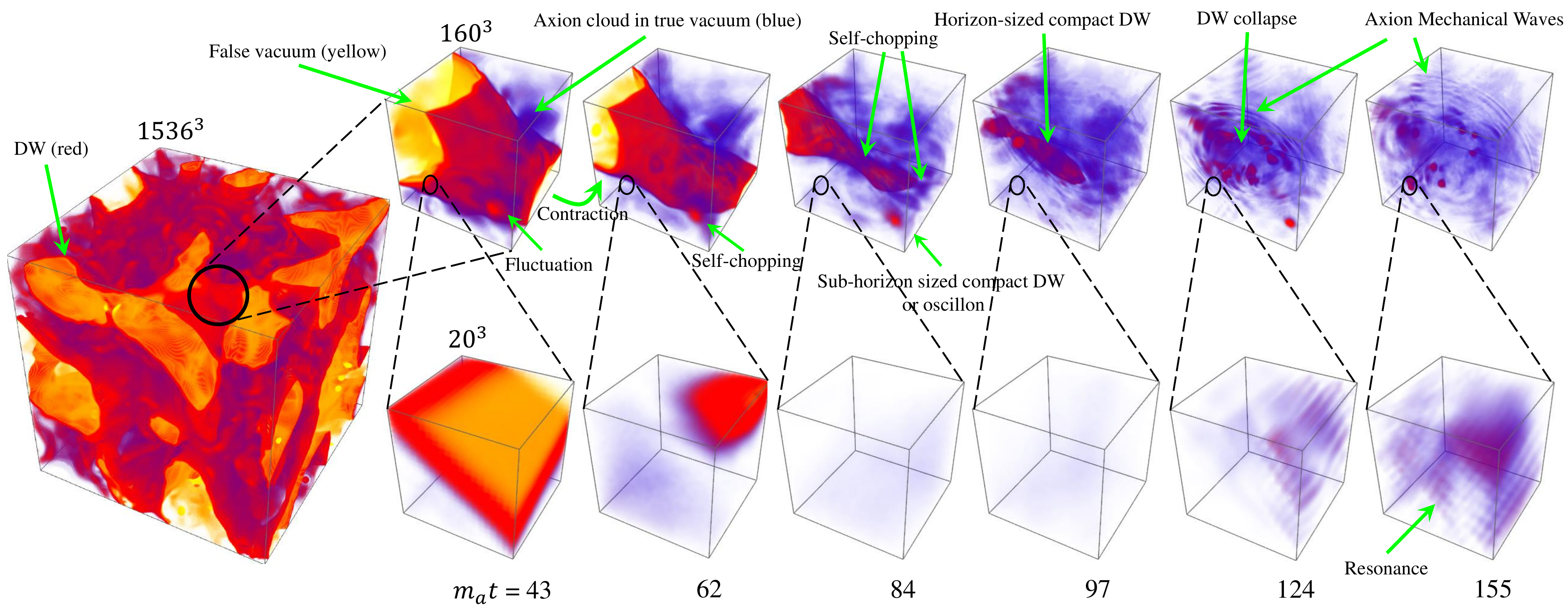}
  \caption{\label{Fig: micro} Visualization of the lattice simulation with the bias parameter $\epsilon = 0.0012$. The leftmost figure displays a snapshot of the entire simulation scale, where the domain wall (DW) is highlighted in red color. The upper row shows a zoomed-in region of our simulation with a $160^3$ lattice, accompanied by a further zoomed-in time series depicted at the bottom. The lower row comprises smaller lattice sizes. Both sets of sub-figures encompass a range of features discovered in this study, and a detailed discussion of these features is provided in the main text of Section~\ref{Sec: Features Identified from Simulation}.}
\end{figure*}

\begin{figure*}[t]
 \includegraphics[width=1.03\textwidth]{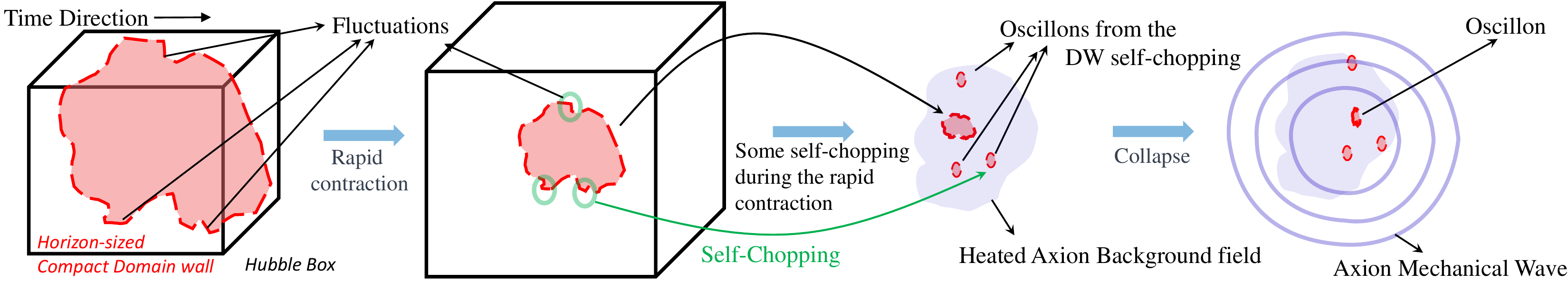} 
  \caption{\label{Fig: Domain_Collapse} A cartoon illustration for the DW collapse process.}
\end{figure*}

\begin{figure*}[t]
 \includegraphics[width=1.03\textwidth]{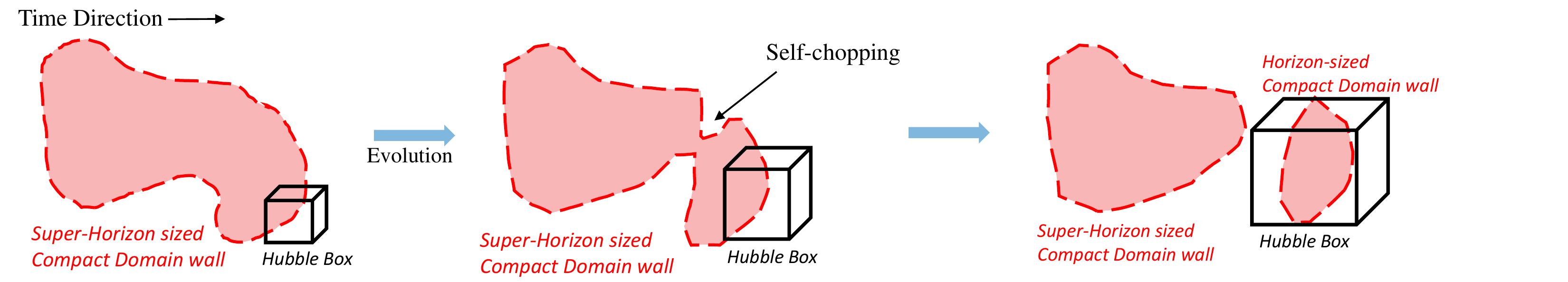} 
  \caption{\label{Fig: Domain_Self-Chopping} A cartoon for DW self-chopping process.}
\end{figure*}

\begin{figure*}[t]
 \includegraphics[width=1.03\textwidth]{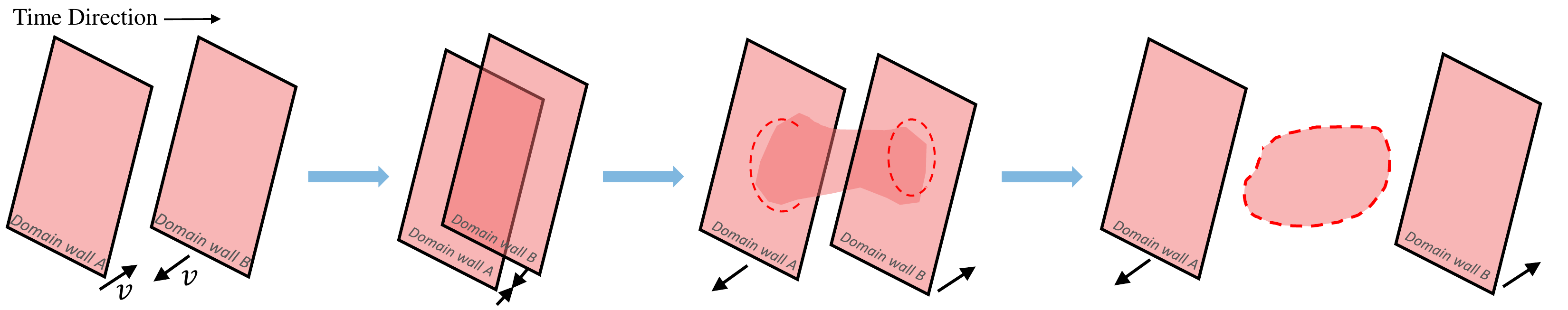} 
  \caption{\label{Fig: Two_Domain_Collision} A cartoon for two DWs chopping. This process is rarely observed in the simulation.}
\end{figure*}

\begin{figure}[t]
 \includegraphics[width=0.5\textwidth]{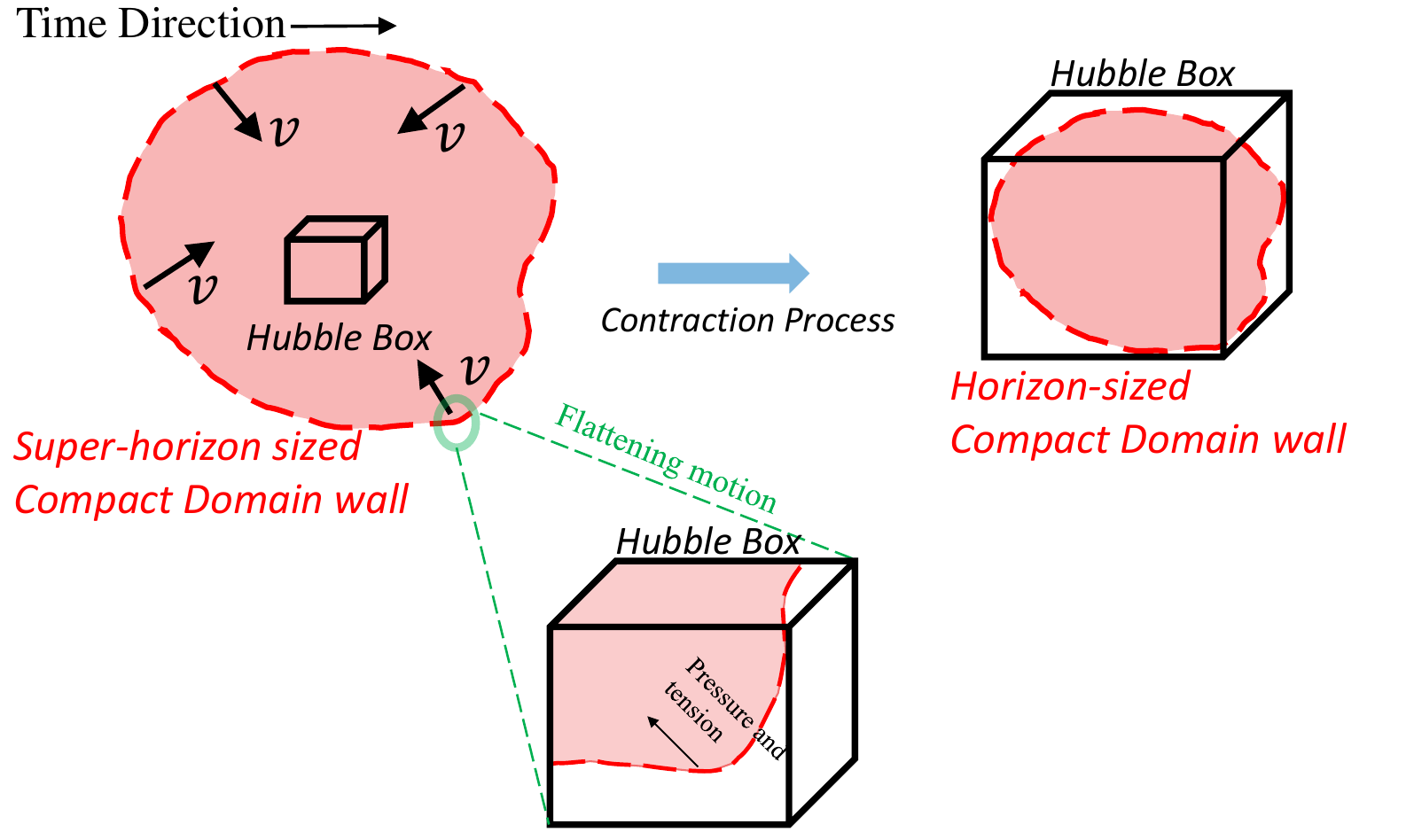} 
  \caption{\label{Fig: Domain_Contraction} A cartoon for DW contraction process with DW velocity $v$. The Hubble box enlarges over time, while the compact DW is contracting. The zoomed-in subfigure illustrates the DW tension and pressure that are also shown in Fig.~\ref{Fig: Domain_Tension}. }
\end{figure}

\begin{figure}[t]
 \includegraphics[width=0.5\textwidth]{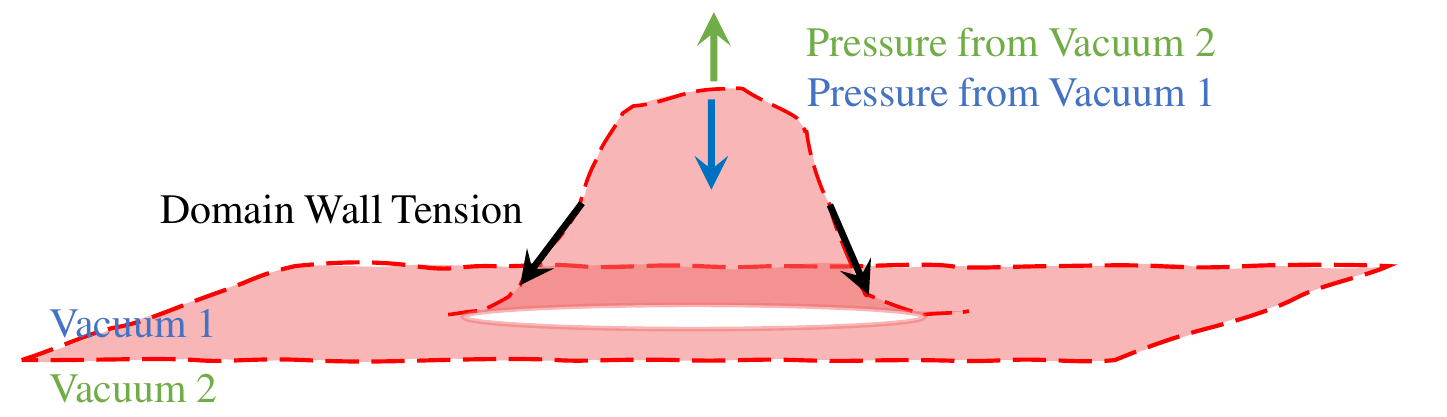} 
  \caption{\label{Fig: Domain_Tension} Force diagram for domain wall tension and vacuum pressures. This is used to illustrate the origin of the DW flattening motion.}
\end{figure}

\subsection{\label{Sec: Features Identified from Simulation} Features Observed in the Simulation}

In this subsection, we will discuss the features identified from the snapshots of our simulations, and will further discuss their corresponding energy contributions and dynamic behaviors later in Sec.~\ref{Free Axion Spectral Analysis} and Sec.~\ref{Model for DW}. We find 6 distinguishable objects in simulation, and they are connected through 3 different dynamic motions, including their creation, annihilation, and motion. 

As illustrated in Fig.~\ref{Fig: micro}, the \textit{objects} observed in the simulation can be categorized as follows:\\
(1) \underline{Super-horizon sized DWs}: represented as the red wall-like structures in Fig.~\ref{Fig: Simulation} and Fig.~\ref{Fig: micro}, with different shapes (either planar or compact). These super-horizon sized DWs are formed due to the initial field distribution of the simulation.\\
(2) \underline{Horizon-sized compact DW}: also shown as red wall-like structures in Fig.~\ref{Fig: Simulation} and Fig.~\ref{Fig: micro}, but with a compact geometry. These horizon-sized compact DWs are formed by the contraction or self-chopping process (which will be discussed) of super-horizon sized DWs. Such DWs release energy through flattening motion, self-chopping into smaller compact DWs, and then collapse (to be defined later). \\
(3) \underline{Sub-horizon compact DWs} or \underline{oscillons}: DWs with typical sizes of $\sim 1/m_a$ (in our simulation it is found that larger, sub-horizon sized compact DW rapidly contract down to the size of $\sim 1/m_a$), much smaller than the horizon scale. These structures are mainly formed through self-chopping due to the fluctuations on the DW surface, and the collapse of the horizon-sized  compact DWs, see Fig.~\ref{Fig: micro} and the red dots in Fig.~\ref{Fig: Simulation}. Distinguishing between sub-horizon compact DWs and oscillons is challenging due to limited lattice resolution, as both structures occupy only a few lattice spacings. Therefore, sub-horizon compact DW and oscillon are two interchangeable terms in this study. At the end of our simulation, sub-horizon compact DWs/oscillons are found to contribute to the residual energy density . However, their contribution is subdominant when compared to that from free axion fields, such as axion clouds and mechanical waves (will be introduced next).\\
(4) \underline{Axion clouds}: background axion field distributed around the vacua, on average with relatively large momentum of $k\gtrsim m_a$. They are shown as blue regions in the true vacuum and yellow regions in the false vacuum in Fig.~\ref{Fig: micro}. 
The formation of axion clouds can be induced by heating the background axion field, i.e. increasing the oscillation amplitude (and thus the energy density) of the background axion field around the vacua through DW movements, specifically, processes like flattening and compact DW collapsing, which will be elaborated on shortly.  \\
(5) \underline{Axion Mechanical Wave}: the ripple-like structure in Fig.~\ref{Fig: micro}, originated from the axion waves propagating outward from the collapsing DWs (see Fig.~\ref{Fig: Domain_Collapse}). Compared to axion clouds, they have relatively lower momentum $k \lesssim m_a$. \\ 
(6) \underline{Resonance}: the phenomenon where a region of the axion clouds are divided into small wave-packets as particle-like structures, with a scale of $k \sim 3.68/m_a$ (this characteristic scale will be demonstrated by spectral analysis later in Sec.~\ref{Free Axion Spectral Analysis}). This characteristic $k$ value is obtained by first visualizing it in spatial dimension and then converting it to momentum space by Fourier transformation. We have shown the resonance in the lowermost-right sub-figure in Fig.~\ref{Fig: micro}.

These objects are connected to each other via transformative processes (e.g. creation, annihilation) which can be categorized as the following \textit{dynamic motions} as identified from our simulation:\\
(1) \textit{Flattening} motion: This DW motion is analogous to laying a piece of paper flat (for example, as illustrated in Fig.~\ref{Fig: Simulation no-epsilon}), and therefore we refer to this motion as ``flattening", which originates from the DW tension and vacuum pressures, see Fig.~\ref{Fig: Domain_Tension}. As a result of such a flattening process, the DW curvature and surface fluctuation are reduced, resulting in the heating of the background axion field. Additionally, the flattening process induces the \textit{contraction} of compact domain walls, causing larger domain walls to transform into smaller ones. For instance, a super-horizon-sized compact domain wall contracts into a horizon-sized domain wall, as illustrated in Fig.~\ref{Fig: Domain_Contraction}.\\
(2) \textit{Self-chopping}: 
refers to the phenomenon where a segment of the DW shrinks and eventually breaks off from the `parent' DW, leading to the splitting of the DW into two parts. This mechanism plays a crucial role in the DW network evolution, affecting any size of DW. The upper-row subfigures in Fig.\ref{Fig: micro} illustrate the sub-horizon sized DW self-chopping (the first three subfigures) and horizon-sized DW self-chopping (the third subfigure) processes, while a cartoon illustration can be found in Fig.~\ref{Fig: Domain_Self-Chopping}. This process is in analogy to self-intersection in cosmic string dynamics. Note that self-chopping is an intermediate process of transforming the DW energy from larger DWs to smaller DWs, which would further decay to the final outcome (mostly free axions) through the collapse process as defined below. Therefore we do not count self-chopping as an effective mechanism of DW energy release, unlike the flattening and collapse processes.\\
(3) \textit{Collapse} of horizon-sized compact DWs: the process during the final stage of DW evolution when a horizon-sized compact DW rapidly contracts and subsequently collapses while radiating the axion field in the form of mechanical wave and heating the background axion field. Such a process is illustrated in the upper-row subfigures in Fig.~\ref{Fig: micro}, and also as a cartoon in Fig.~\ref{Fig: Domain_Collapse}. 

The complete evolution process of DWs can then be summarized as follows: At the beginning of simulation, super-horizon-sized DWs transform into horizon-sized DWs via either contracting (by flattening) or dividing (by self-chopping). Following this, the horizon-sized DWs undergo a collapse, resulting in the emergence of axion mechanical waves and axion clouds, while also releasing a smaller amount of energy in form of subhorizon compact DWs. Throughout this entire process, the sub-horizon-sized DWs and oscillons undergo continuous self-chopping, while the background axion field continues to heat up, resulting in the formation of an axion cloud.

The energy released during the evolution of DWs can be categorized based on two key mechanisms: flattening and collapsing. In Section \ref{Model for DW}, we will delve into a detailed discussion and analysis of the energy release, with a specific emphasis on these two aspects i.e.~collapsing, leading to $\rho_2$ in Eq.(\ref{Eq: EoM rho_2}) and flattening, leading to $\rho_3$ in Eq.(\ref{Eq: EoM rho_3}), respectively. Note that it may not be feasible to precisely separate the energy contributions arising from these two mechanisms, as both of them lead to the heating of the background axion field. There are additional contributions from processes such as the self-chopping and the subsequent decay of sub-horizon compact DWs. But these influences are comparatively insignificant when compared to the essential processes mentioned above.

It is worth noting that in the analogous VOS model of cosmic strings, the majority of energy is released through the formation of loops primarily generated by the interaction of two long strings \cite{Kibble:1984hp}. In contrast to the chopping process of cosmic strings, the probability of chopping due to the intersection of two DWs (cartoon illustration in Fig.~\ref{Fig: Two_Domain_Collision}) is negligibly low, and the majority of energy loss is due to the two mechanisms-flattening and collapse outlined above. The energy contribution from the self-chopping of sub-horizon compact DWs is negligible when compared to that of horizon-sized compact DWs, as found in the simulation. Furthermore, it is observed that horizon-sized and sub-horizon sized compact DWs typically do not originate from the chopping of two horizon-sized DW, as shown in Fig.~\ref{Fig: Two_Domain_Collision}, rather, from contraction or self-chopping of larger, super-horizon or horizon-sized DWs. 

\subsection{\label{Sec: Scaling behavior}Scaling Regime}

In our simulation, we track the evolution of DWs and the pattern of energy loss from the DW network. A snapshot of the evolution is shown in Fig.~\ref{Fig: Simulation}, and for comparison, its counterpart with non-biased potential is shown in Fig.~\ref{Fig: Simulation no-epsilon} in the Appendix.\ref{Appendix}. The left-most snapshot is taken as the network enters the scaling regime when the DWs flatten while expanding. Shortly after its formation ($\Delta t \lesssim 10/m_a$), the network approaches an attractive solution called the scaling regime while releasing energy through the two mechanisms which were introduced in the last section: (1) the collapsing of horizon-sized compact DWs; (2) The flattening motion. Meanwhile, the super-horizon DWs enter into the horizon continuously, which consequently compensates for the energy loss due to both mechanisms, so that the DW area per horizon volume $A_v$ remains constant. This constant solution is the feature of the scaling regime. Such a feature has been identified in literature \cite{Avelino:2005kn,Martins:2016lzc,Kawasaki:2013ae,Kawasaki:2014sqa,Hiramatsu:2013qaa}, and also agrees with our findings as shown in Fig.~\ref{Fig: AVt_mat}. At a later time, the DWs start to decay around $t_{\rm decay}$, and the scaling solution breaks down. In the scaling regime, the DW energy density takes the following form:
\begin{align}
\label{Eq: DW energy density}
    \rho_{\rm DW} = \gamma^2 \frac{\sigma_{\rm DW} A_v}{t},
\end{align}
where $\gamma \simeq 1$ is the Lorentz factor that represents the contribution of the kinetic energy of the DW, and the DW area parameter is given by (originally introduced in \cite{Press:1989yh})
\begin{align}
\label{Eq: A_v}
A_v \equiv \frac{A_w t}{R(t) V} = 0.67^{+0.04}_{-0.04}, \;\;\;\;\;\;\;\hbox{for} \;\;\epsilon = 0,
\end{align} 
where $A_w$ is the DW comoving area, and $V$ is the comoving volume. The result largely agrees with the previous simulation studies \cite{Kawasaki:2014sqa,Hiramatsu:2013qaa,Hiramatsu:2012sc,Hiramatsu:2013qaa}, but it is about $30\%$ less than the prediction by the simulation assuming PRS approximation for the DW network \cite{Martins:2016ois}. On the other hand, in the metastable DW scenario, we find
\begin{align}
\label{Eq: A_v Fitting Model}
   A_v =   c_1 + c_2 \, \hbox{Exp}\left[-c_3\, \left(\epsilon \sqrt{m_at}\right)^{c_4}  \right], \;\;\;\;\hbox{for} \;\; \epsilon > 0 ,
\end{align}
with 
\begin{align}
 \notag &  c_1 = 0.0088^{+0.0009}_{-0.0009}, \;\;\;
        c_2 = 0.62^{+0.06}_{-0.05}, \\  \notag 
         &   c_3 = 3.98^{+0.40}_{-0.40}  \times 10^{6}, \;\;\;
                c_4 = 3.57^{+0.08}_{-0.11}, 
\end{align}
where the parameter $c_1$ term represents the residual compact DWs and oscillons at the end of the simulation. As mentioned we cannot distinguish whether these are sub-horizon (i.e. much smaller than the horizon scale) compact DWs or oscillons due to the limitation of the simulation period and resolution. The fitting model Eq.(\ref{Eq: A_v Fitting Model}) is inspired by field theory analysis \cite{Hindmarsh:1996xv} that employs mean-field approximation method and Gaussian ansatz on the field probability distribution in the limit of a small bias term $\epsilon \ll 1$. Moreover, the parameter $c_4 \sim 3$ is approximately the spatial dimension as predicted in \cite{Hindmarsh:1996xv}. The fitting model in Eq.(\ref{Eq: A_v Fitting Model}) also fits the data from other DW simulation studies \cite{Larsson:1996sp,Correia:2014kqa,Correia:2018tty}. As the axion kinetic energy reduces due to redshifting, the true vacuum pressure force gradually overcomes the DW tension, which causes energy loss of the DW network. We define the characteristic decay time of the DW, $t_{\rm decay}$, as when the DW area $A_v$ becomes $\sim 10\%$ of the pre-collapsing value i.e.~$ 0.1 A_v(t\to 0) = A_v(t_{\rm decay})$. $t_{\rm decay}$ can be estimated by Eq.(\ref{Eq: A_v Fitting Model}) as 
\begin{align}
\label{Eq: t_decay}
    t_{\rm decay} & \simeq \frac{\epsilon^{-2}}{m_a} \left(\frac{c_\mu}{c_3}\right)^{2/c_4} \\ \notag &  \simeq\frac{\epsilon^{-2}}{m_a}  (3.22 \pm 0.94) \times 10^{-4} ,
\end{align}
where the factor 
\begin{align}
    c_\mu = 2.32_{-0.60}^{+0.61}.
\end{align} 
Note that other semi-analytical estimation studies \cite{Kawasaki:2014sqa,Saikawa:2017hiv} compare the pressure gap between vacua and use a power-law model to fit their data, and predict $t_{\rm decay} \propto 1/\epsilon$. This causes a notable difference from our results in the prediction for the axion relic abundance as shown in Sec.~\ref{Sec: Cosmological implication}.  

\subsection{\label{Sec: DW velocity}Domain Wall Velocity}  

\begin{figure}[t]
 \includegraphics[width=0.49\textwidth]{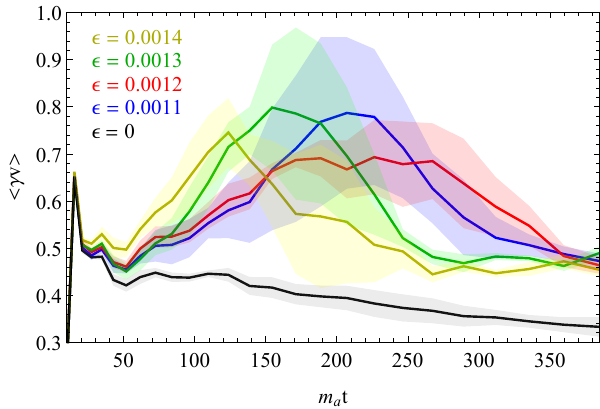}
  \caption{\label{Fig: Velocity} The average $\gamma v$ versus $m_a t$ with varying bias parameter $\epsilon$. The uncertainty bands are shown as shaded areas. The network enters the scaling regime at about $m_a t \sim 15$, thus the earlier peak at $m_a t \sim 10$ is due to the initial condition.} 
\end{figure}

\begin{figure}[t]
 \includegraphics[width=0.49\textwidth]{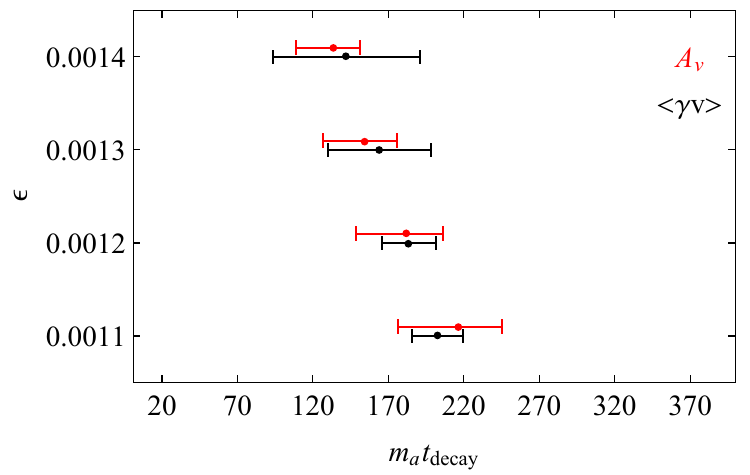}
  \caption{\label{Fig: Velocity_peak} Bias parameter $\epsilon$ versus axion mass times decaying time $m_a t_{\rm decay}$. The red bars are the decay time calculated by Eq.(\ref{Eq: t_decay}) using the fitting result of this study with Eq.(\ref{Eq: A_v Fitting Model}). The black bars are estimated from the peaks in Fig.~\ref{Fig: Velocity}, which is the time that DW velocity starts deceleration. }
\end{figure}

In DW dynamics, its velocity plays an important role in its equation of motion. We measure the velocity by tracking the movement of the maximum of the axion potential $V(a(x,t)) = V_{\rm max}$ in the simulation. The observed DW velocity is shown in Fig.~\ref{Fig: Velocity} for varying $\epsilon$. The DW network during the scaling regime at first decelerates (relative to the initial velocity set by initial condition) due to the Hubble friction and the DW flattening motion, then experiences acceleration due to the pressure difference between the true and false vacua, during the decaying period $t \sim t_{\rm decay}$, then decelerates again when the network decays away during the later stage of $t > t_{\rm decay}$. The peak of each curve is thus located at about $t_{\rm decay}$, see Fig.~\ref{Fig: Velocity_peak}, where we show that the comparison of the decay time $t_{\rm decay}$ as defined in Eq.(\ref{Eq: t_decay}) and the peak of the observed velocity. 

To fit the DW velocity function, we consider the following model:
\begin{align}
\label{Eq: gamma velocity}
    \gamma v  = \frac{0.923 \pm 0.136}{(m_a t)^{0.614\pm 0.031}} + \alpha_v e^{-( t - t_{\rm decay} )^2/(2\sigma_v^2)},
\end{align}
with
\begin{align}
\notag  \alpha_v = (0.241 \pm 0.039), \;\;\; \hbox{and} \;\;\;   \sigma_v = (52\pm 20) \frac{1}{m_a} .
\end{align} 
The second term in Eq.(\ref{Eq: gamma velocity}) indicates the effect of the pressure difference between the true and false vacua in the decay phase, $\alpha_v$ represents the magnitude of the acceleration, $\sigma_v$ is the uncertainty in our observation and the exponential indicates that the acceleration stops at about $t\simeq t_{\rm decay}$.

This section analyzed the the domain wall's velocity, which along with earlier discussions, paves the way for the next section, where we will investigate the free axion spectrum, resulting from of the decay of the DWs.

\section{\label{Free Axion Spectral Analysis}Free Axion Spectral Analysis}

\begin{figure}[t]
\includegraphics[width=0.485\textwidth]{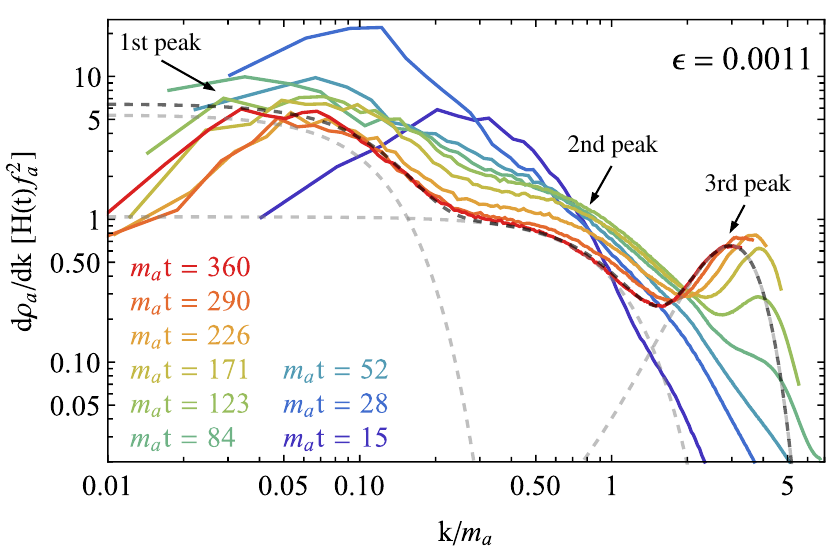} 
\caption{\label{Fig: PRho-k 11} Free axion energy density spectrum $\partial \rho_a / \partial k $ as a function of physical momentum $k$, assuming the bias parameter $\epsilon = 0.0011$. The early to later spectrum is shown as blue to red. The spectrum can be split into three Gaussian distributions as shown as dashed gray curves corresponding to the 3 contributing terms in Eq.(\ref{Eq: Fitting_model}). From low $k$ to higher $k$, these three Gaussian distributions present the energy density from misalignment ($k/m_a \lesssim 0.2$), free axions radiated by compact DW self-chopping, and collapsing ($k/m_a \lesssim 1$), and the small structure axion field such as the axion clouds with the resonance at ($k \sim \mathcal{O}(m_a)$ ), respectively. The smaller $k < 0.01 m_a$ region is lacking data because of the simulation lattice size, and higher $k$ has been cut at Nyquist frequency as discussed in Sec.~\ref{Free Axion Spectral Analysis}.
}
\end{figure}
We discuss the details of the spectral analysis for free axion energy density in this section, which would be the key input for estimating axion dark matter relic density in Sec.~\ref{Sec: Cosmological implication}. As discussed in Sec.~\ref{Sec: Setup}, we estimate the total free axion energy as twice the masked axion kinetic energy. We then compute the free axion spectrum according to \cite{Gorghetto:2018myk,Hiramatsu:2012sc} as
\begin{align}
\label{Eq: rho_a_with_kinetic}
    \rho_a = \int dk \partial \rho_a / \partial k, \;\;\;\; \hbox{with}  \;\;\;\;  \rho_a = \langle \dot{a}^2 \rangle,
\end{align}
where the axion spectrum $\partial \rho_a / \partial k$ is given by
\begin{align}
    \frac{\partial \rho_a}{\partial k} = \frac{k^2}{(2 \pi L)^3} \int d \Omega_k |\tilde{\dot{a}}(k)|^2,
\end{align}
where $\tilde{\dot{a}}(k)$ is the Fourier transform of $\dot{a}(x)$, $L = (N/2) R(t)\Delta x_i$ is the collected data range, and the momentum $\vec{k} \equiv \frac{2\pi \Vec{n}}{L}$. In addition, we cut off the momentum that is higher than the Nyquist frequency $k_{\rm Ny} = \pi/(R(t)\Delta x_i)$ to prevent corrupted data caused by insufficient resolution. 

In Fig.~\ref{Fig: PRho-k 11} we show the free axion energy spectrum with snapshots for the cosmic time evolution, using different colors. The dark blue curve ($m_a t = 15$) represents the spectrum when the network just enters the scaling regime, and the red curve ($m_a t = 360$) presents the spectrum near the end of the simulation. We find that the spectrum can be fitted as a sum of three Gaussian distributions corresponding to distinct physics origins (to be explained later):
\begin{align}
\label{Eq: Fitting_model}
   \frac{\partial \rho_a}{ \partial k} &= \sum_{i=1}^3 \frac{\partial \rho_i(A_i, k_i, \sigma_i)}{\partial k},
\end{align}
where the labels $i=\{1,2,3\}$ denote the 3 gray-dashed curves from low k to higher in Fig.~\ref{Fig: PRho-k 11}, associated with the first, second, and third peak, as indicated respectively. These curves are parameterized by
\begin{align}
    \frac{\partial \rho_i(A_i, k_i, \sigma_i)}{\partial k} \equiv A_i \exp^{-(k-k_i)^2/\sigma_i^2},
\end{align}
where we set $k_1 \simeq 0$ due to the lack of data within the large scale range of $k \leq 0.02$ as limited by the simulation size of $N=1536$, and $k_2 \simeq 0$ since the first peak dominates over the lower $k$ range associated with the second peak, making it challenging to discern the contribution of the second peak to the measurement, and
\begin{align}
\label{Eq: k_3}
    k_{3} = (3.68 \pm 0.03)m_a,
\end{align}
which decreases as $1/R(t)$ due to redshift after DW decay. We fit the parameters in Eq.(\ref{Eq: Fitting_model}) with data from each cosmic time snapshot from every simulation run (we show data from a single run in Fig.~\ref{Fig: PRho-k 11} as an example), then analyze their time dependence in the next section. The fitting results for the parameters and energy densities in Eq.(\ref{Eq: Fitting_model}) are given in Appendix.~\ref{Appendix}.  We have verified the robustness of the peak at $k_3$ by conducting additional test runs involving variations in the value of $m_a$ and lattice spacing, as outlined in Sec.~\ref{Sec: Simulation Setup}, but the magnitude of the peak may be subject to the inherent resolution of the simulation during the later stage of the simulation (roughly when $t\gtrsim300/m_a$), as $k_3$ in Eq.(\ref{Eq: k_3}) closely approaches the Nyquist frequency during this later stage.

We observe that $\rho_1$ is in reasonable agreement with the energy density of axions produced through the misalignment mechanism, specifically, $\rho_1 \sim m_a^2 f_a^2/2N_{\rm DW}^2$ \footnote{Note that the initial condition that sets the axion fields on vacuums seems to exclude the axion energy from the misalignment mechanism. However, it just stores the energy on the gradient energy budget at the onset of simulation.}, at the early stage of simulation, then redshifts like matter. As a result of this redshift, the spectral line associated with this contribution progressively shifts towards the lower frequencies over time. 

The free axion energy density $\rho_2$ in Eq.(\ref{Eq: rho_a_with_kinetic}) carries the energy contribution with the scale $k\lesssim m_a$. We attribute this energy component to axion mechanical wave originated from collapsing of horizon-sized compact DWs.
There are two reasons for this explanation:
(1) The energy spectrum of $\rho_2$ is consistent with the scale range of the axion mechanical wave, i.e. $k\lesssim m_a$.
(2) $\rho_2$ aligns well with the production process of the compact DW according to the data fitting (see Eq.(\ref{Eq: Energy Loss Rho_2}, and the details will be provided in the next section), as predicted by the DW VOS model in the context of DW chopping \cite{Avelino:2005kn}. It is important to note that while we observe the self-chopping phenomenon (as discussed in Sec.~\ref{Sec: Features Identified from Simulation}), it differs from the definition of two DWs chopping in the VOS model. Nonetheless, they share a similar energy loss form in the equation of motion, as we will see in Sec.~\ref{Model for DW}.

The energy density $\rho_3$ can be interpreted as the contribution from axion clouds with a resonance at $k_3$. This energy arises from various processes, as discussed in Sec.~\ref{Sec: Features Identified from Simulation}. We anticipate that the primary contribution to this energy comes from the annihilation of fluctuations on the DW surface through the flattening motion, because the estimation of the energy released from these fluctuations aligns well with the energy density $\rho_3$ as demonstrated in Sec.~\ref{Model for DW}.

The energy release mechanisms discussed in Sec.~\ref{Sec: Features Identified from Simulation} occur in both the scaling regime and decaying period, and the compact DW collapse is more likely to occur in the decaying period. In other words, the biased potential significantly accelerates the DW flattening, contraction, and self-chopping.
During the decaying phase, we find that the production of axion clouds ($\rho_3$) increases by about $\sim 70\%$, and the radiation for larger wavelength axion mechanical waves ($\rho_2$) is enhanced by about $\sim 30\%$, compared to the scaling regime. The percentage is estimated at the time $t_{\rm decay}(\epsilon\to 0.0012)$, and by comparing the outcome from the $\epsilon = 0.0012$ and $\epsilon = 0$ scenarios. 

\section{\label{Model for DW}Model for Domain Wall Evolution} 

In this section, we present the coupled evolution equations for the energy densities of the DW network and of the free axions emitted from the DWs. The two components of axion energy densities sourced by different DW dynamics, $\rho_2$ and $\rho_3$, as identified via spectral analysis and monitoring simulation evolution in Sec.~\ref{Sec: Features Identified from Simulation} and Sec.~\ref{Free Axion Spectral Analysis}, are key inputs in this section. Here we will quantitatively model these contributions, $\rho_2$ and $\rho_3$, respectively, by numerically fitting simulation data. We extract time-dependent data from simulations in Sec.~\ref{Free Axion Spectral Analysis}, and we further fit them into the DW evolution equations in this section. 

We first generalize the DW evolution equation in the VOS model for a stable DW network \cite{Avelino:2005kn, Martins:2016lzc} as follows:
\begin{align}
\label{Eq: EoM DW}
    \frac{d \rho_{\rm DW}}{dt} = - (1 + 3 v^2) H \rho_{\rm DW} - \frac{d \rho_{\rm DW}}{dt}\Bigg|_{\rm to 2} - \frac{d \rho_{\rm DW}}{dt}\Bigg|_{\rm to 3},
\end{align}
where the right-hand side of the equation represents, in order, the redshift effect, the DW energy loss to $\rho_2$ and $\rho_3$, respectively.
Here we have reasonably assumed that the final form of DW energy release is free axions, as gravitational wave radiation albeit inevitable, is expected to be subleading. 

\begin{figure}[t]
\includegraphics[width=0.45\textwidth]{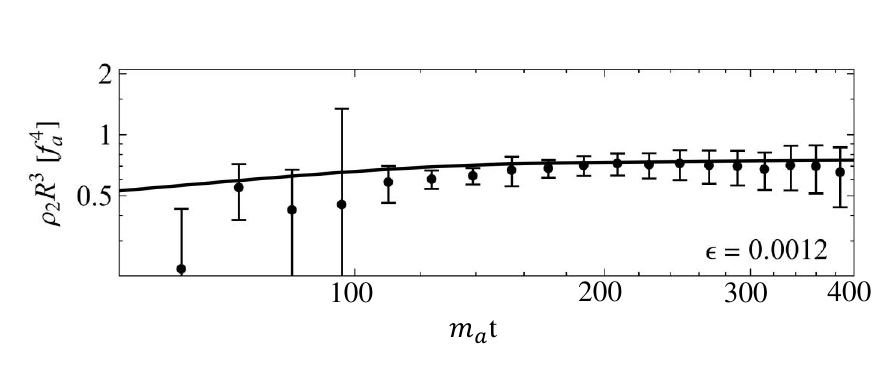} 
\caption{\label{Fig: Fitting epsilon peak 2 11} The energy density of the second Gaussian fitting function (related to $\rho_2$) as given in Fig.~\ref{Fig: PRho-k 11} and Eq.(\ref{Eq: Fitting_model}) where we fix $\epsilon = 0.0012$. The black curve presents the prediction of axion production model Eq.(\ref{Eq: EoM rho_2}) which implies the energy loss in the DW network through the horizon compact DW collapse as discussed in Sec.~\ref{Sec: Features Identified from Simulation}. }
\end{figure}

By energy conservation, the latter two terms in Eq.~\ref{Eq: EoM DW} also enter the evolution equations of the free axions, which is essential for solving the axion relic abundance. As revealed via the spectral analysis based on simulation results, free axion production from DWs can be roughly divided into two kinetic regions associated with distinct DW dynamics, corresponding to $\rho_2$, $\rho_3$. It is thus reasonable to consider the evolution of $\rho_2$ and $\rho_3$ components separately, then sum up their solution for the total axion abundance. We first write down the evolution equation for $\rho_2$, which originates from the collapse of compact DWs:
\begin{align}
\label{Eq: EoM rho_2}
    \frac{d\rho_2}{dt} = - 3 H \rho_2 + \frac{d\rho_{\rm DW}}{dt}\Bigg|_{\rm to 2 },
\end{align}
where $3H$ reflects the finding that this spectral component of axions generally has a longer wavelength and behaves like cold matter, and corresponds to the axion mechanical wave as introduced earlier. 
The second term on the right-hand side reflects energy conservation and the aforementioned reasonable assumption that the DW energy release $100\%$ goes to axions. As the second term descends from the formation of compact DWs through DW self-chopping, we can explicitly model its evolution as follows:
\begin{align}
\label{Eq: Energy Loss Rho_2}
    \frac{d \rho_{\rm DW}}{dt}\Bigg|_{\rm to 2} = \tilde{c}_v v  \frac{\rho_{\rm DW}}{L_{\rm DW}},
\end{align}
where the self-chopping efficiency parameter, $ \tilde{c}_v$, can be modeled as
\begin{align}
\label{Eq: c_v tilde}
       \tilde{c}_v \equiv c_v  \gamma^{c_\gamma} \mathcal{A}_F^{-c_{\mathcal{A}}}, 
\end{align}
with
\begin{align}
    c_v = 0.36^{+0.07}_{-0.03}, \;\;\;\; c_\gamma = 3.36^{+0.93}_{-0.58}, \;\;\;\; c_{\mathcal{A}} = 1.55^{+0.04}_{-0.06}
\end{align}
where $L_{\rm DW} = \gamma^2 \sigma_{\rm DW}/\rho_{\rm DW}$ is the DW correlation length.
The value of the parameters $c_v$, $c_\gamma$, and $c_{\mathcal{A}}$ are calibrated by the simulation data. A single run data is shown in Fig.~\ref{Fig: PRho-k 11} and Fig.~\ref{Fig: Fitting epsilon peak 2 11}. $\mathcal{A}_{F}$ is the area fraction parameter:
\begin{align}
    \mathcal{A}_{F} \equiv \frac{A_v(\epsilon)}{A_v(\epsilon \to 0)}.
\end{align}
where $A_v$ is defined in Eq.(\ref{Eq: A_v Fitting Model}). In the limit of non-relativistic and stable DW, i.e., $\gamma \to 1$ and $\epsilon \to 0$, Eq.(\ref{Eq: Energy Loss Rho_2}) approaches the expression $c_v v \frac{\rho_{\rm DW}}{L_{\rm DW}}$ which was used to describe the energy loss resulting from the intersection of DWs,  leading to the creation of compact DWs that eventually collapse. This term was originally introduced by Kibble in the context of the cosmic string network \cite{Kibble:1984hp}, and later applied to the stable DW VOS model \cite{Avelino:2005kn} for two DWs chopping. We slightly modify its physical interpretation to self-chopping and utilize it to explain our data (see Fig.~\ref{Fig: Fitting epsilon peak 2 11}). The factor $\mathcal{A}_F^{-3/2}$ captures the simulation finding that compact DW production is more efficient during the decay phase, $v \rho_{\rm DW}/L_{\rm DW}$ represents the likelihood of DW self-chopping, and $\gamma^{c_\gamma}$ indicates that an accelerated DW velocity increases the rate of self-chopping.

We further estimate the solution of $\rho_2$ by numerically solving the axion radiation equation Eq.(\ref{Eq: EoM rho_2}) with Eq.(\ref{Eq: Energy Loss Rho_2}), and can be fitted as:
\begin{align}
\label{Eq: app rho_2}
\rho_2 \left(\frac{R(t)}{R(t_{\rm decay})}\right)^3 \simeq 2 \tilde{c}_v v \rho_{\rm DW} \bigg|_{\epsilon \to 0, \; t \to t_{\rm decay}}.
\end{align}
The dominant DW contribution to the $\rho_2$ component of the axions is from the era around $t_{\rm decay}$, and the radiated axions redshift like matter afterward. This solution can be understood as resulting from energy conservation. 

\begin{figure}[t]
\includegraphics[width=0.45\textwidth]{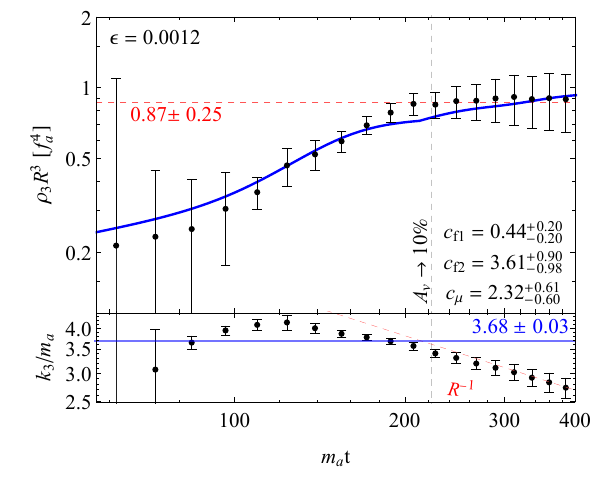} 
\caption{\label{Fig: Fitting epsilon 12} The energy density for the third Gaussian fitting function (related to $\rho_3$) as given in Fig.~\ref{Fig: PRho-k 11} and Eq.(\ref{Eq: Fitting_model}), with $\epsilon = 0.0012$. The blue curve presents the prediction of axion production model Eq.(\ref{Eq: EoM rho_3}) which implies the energy loss in the DW network through the DW flattening motionas discussed in Sec.~\ref{Sec: Features Identified from Simulation}. The vertical line $A_v \to 10\%$ corresponds to the time $t_{\rm decay}$ when $A_v$ becomes $10\%$ of the value at scaling regime (Eq.~\ref{Eq: A_v}).}
\end{figure}

Next, we consider the evolution equation for the component of $\rho_3$, mostly due to the axion clouds production from the  DWs flattening motion as discussed in Sec.~\ref{Sec: Features Identified from Simulation}. By analogy of Eq.~\ref{Eq: EoM rho_2} for $\rho_2$, we have:
\begin{align}
\label{Eq: EoM rho_3}
    \frac{d\rho_3}{dt} = - \lambda_3 H \rho_3 + \frac{d\rho_{\rm DW}}{dt}\Bigg|_{\rm to3},
\end{align}
where $\lambda_3$ represents the time-dependent redshift of this component of axion energy density. As shown in the spectral analysis, at production these axions are on average (semi-)relativistic with a shorter wavelength, thus radiation-like and $\lambda_3 \simeq 4$; then the axions cool down and become matter-like with $\lambda_3 = 3$
\footnote{The emitted axions can be thought of as hot axions at first. Our simulations have confirmed that when the initial conditions of the axion field are such that the time derivative $\dot{\theta} \gg m_a$ and the spatial gradient $\nabla_x \theta \gg m_a$ (with $\theta = a(x)/f_a$). This means that during the early stage the kinetic and gradient components dominate over the potential energy $V(a) \sim m_a^2f_a^2$. In this scenario, the axion energy at first oscillates harmonically between the kinetic and gradient components, when the total energy density dilutes like radiation. As the kinetic energy later becomes comparable to the potential energy, the axion rolls down to the potential minimum and starts exhibiting characteristics as a matter-like component.}. For simplicity, we use the following function for $\lambda_3$ to fit the spectrum,
\begin{align}
    \lambda_3 = \left\{           
\begin{aligned}
\, 4 \;\;\;\; \hbox{for} \;\;\;\; t< t_{\rm decay}, \\
\, 3 \;\;\;\; \hbox{for} \;\;\;\; t\geq t_{\rm decay}.
\end{aligned}
\right.
\end{align}
The evolution of DW energy loss that leads to this component of axion production can be modeled as (to be explained later):
\begin{align}
\label{Eq: Energy Loss Rho_3}
\notag    \frac{d \rho_{\rm DW}}{dt}\Bigg|_{\rm to 3} = & \; \frac{1}{2} \frac{d}{dt} \left[  \rho_{\rm DW} (1 - v^2)^{c_{f2}} \left( \frac{m_a}{H} \right)^{c_{f1} (1 - \mathcal{A}_F )} \right],\\
    \equiv & \; \frac{1}{2} \frac{d}{dt} \mathcal{F}_A(t),
\end{align}
where the parameters are calibrated by simulation data as: 
\begin{align}
\label{Eq: cf1 and cf2}
    c_{f1} = 0.44^{+0.20}_{-0.20}, \;\;\;\; c_{f2} = 3.61^{+0.90}_{-0.98}.
\end{align} 
We also show a fitting result for $\epsilon = 0.0012$ in Fig.~\ref{Fig: Fitting epsilon 12} as an example. Similar to the case of $\rho_2$, the numerical solution of Eq.~(\ref{Eq: EoM rho_3}) can be fitted as
\begin{align}
\label{Eq: app rho_3}
    \rho_3 \left(\frac{R(t)}{R(t_{\rm decay})}\right)^3 \simeq \mathcal{F}_A\bigg|_{\epsilon \to 0, \; t \to t_{\rm decay}}.
\end{align}
We have chosen the model fitting form given by Eq.~(\ref{Eq: Energy Loss Rho_3}) for the following reasons.  Firstly, the energy of the perturbation per unit area of the DW increases with the scalar (axion) mass $m_a$ as estimated in \cite{Vachaspati:1984yi}. Additionally, the total area of the horizon-sized DWs within a horizon decreases as $H$ increases, and it is expected that the energy loss of DWs is greater for higher overall DW energy density $\rho_{\rm DW}$. These considerations are captured by the variables $1/H$, and $\rho_{\rm DW}$, respectively, along with their functional form in Eq. (\ref{Eq: Energy Loss Rho_3}). In addition, the power of $c_{f1}(1-\mathcal{A}_F)$ renders the dimensionless parameter $m_a/H$ negligible in the scaling regime, which captures the fact that the DW fluctuations release energy becomes more significant in the scenario of metastable DWs (i.e.~$\epsilon \neq 0$).
We also introduced a simple velocity dependence to Eq.~ (\ref{Eq: Energy Loss Rho_3}) as preferred by numerical fitting, which implies that a significant contribution to $\rho_3$ occurs around the peaks shown in Fig.~\ref{Fig: Velocity}, i.e.~when $t\sim t_{\rm decay}$. 

It is important to note that the DW fluctuation (scalar perturbation) radiation term as described in \cite{Martins:2016ois} represents the axion radiation resulting from the annihilation of surface fluctuations, which corresponds to $\rho_3$ component in this study. They find that the chopping effect \footnote{As mentioned in Sec.~\ref{Sec: Features Identified from Simulation}, the word 'chopping' is used in \cite{Martins:2016ois} for two DW chopping, but we use 'self-chopping' which was defined in the text and found to dominate over two DW chopping from our simulation.} in the VOS model that results in $\rho_2$ in this study is negligible in their simulation. Their conclusion does not align well with the axion spectrum depicted in Fig.~\ref{Fig: PRho-k 11} as found from our simulation. This discrepancy may be attributed to the utilization of the PRS algorithm \cite{Press:1989yh} in \cite{Martins:2016ois}, which can inaccurately model the DW dynamics at small-scale structures, as pointed out in \cite{Hiramatsu:2013qaa}. 

There are also caveats identified from our detailed analysis that are worth reiterating. Firstly, $\rho_3$ encompasses not only the radiation from the flattening the surface fluctuations of the DWs, but also the (sub-dominant) contributions from, for instance, the collapse of horizon-sized compact DW that also leads to the heating of background axion field, as discussed in Sec.~\ref{Sec: Features Identified from Simulation}. Secondly, in the later stages of the simulation, the characteristic energy scale of $\rho_3$ become close to the Nyquist frequency, which may result in considerable observational uncertainties, as discussed in Sec.~\ref{Free Axion Spectral Analysis}. 

In this section we introduced the coupled evolution equations for DW network and free axions from DWs, using $\rho_2$ and $\rho_3$ from spectral analysis, and provided an estimate for axion production. The DW evolution equation, considering the redshift effects and energy loss to $\rho_2$ and $\rho_3$, demonstrates the relation between DW energy loss and axion production. Separate equations for $\rho_2$ and $\rho_3$ capture the horizon compact DW creation and collapse, and axion cloud production and axion field resonance, respectively. In the next section, we will apply the results obtained here for the prediction of $\Omega_a$.

\section{\label{Sec: Cosmological implication}Cosmological implication}

\begin{figure}[t]
\includegraphics[width=0.48\textwidth]{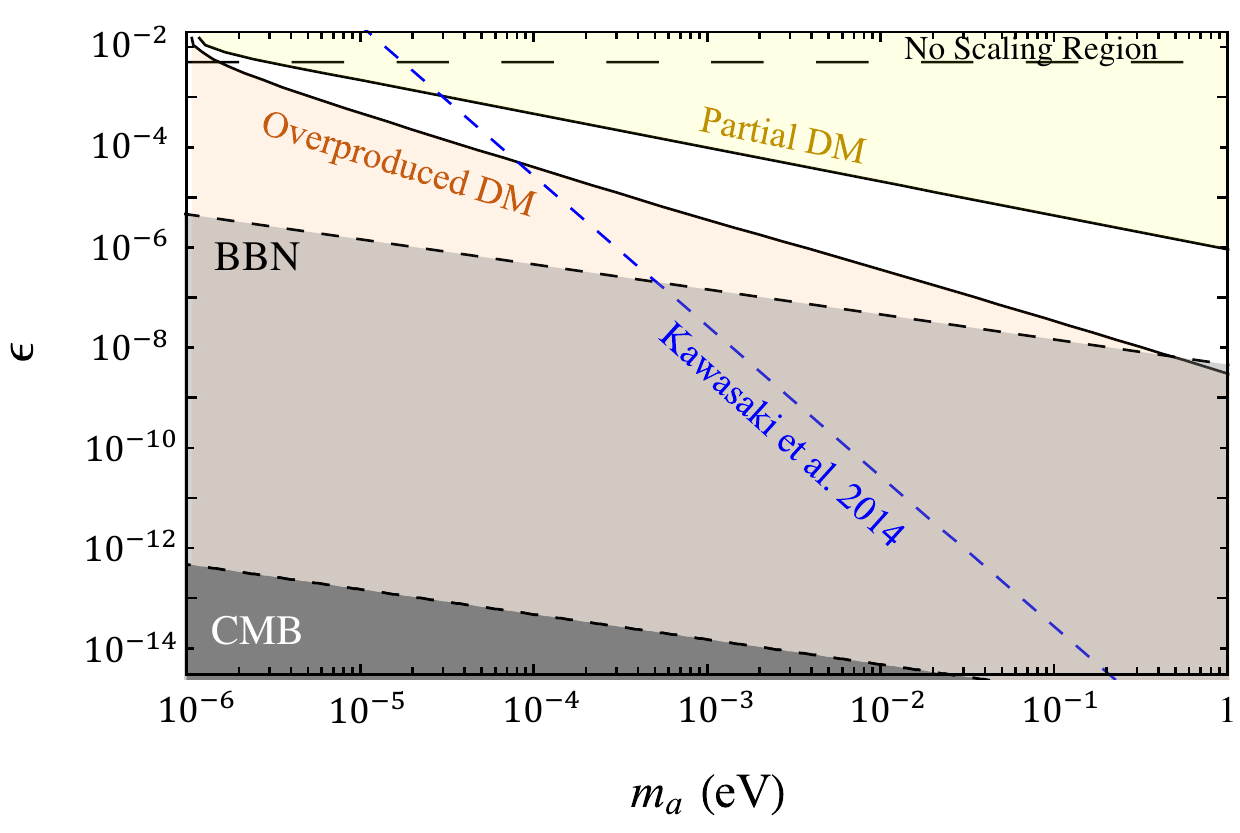} 
\caption{\label{Relic_NoString_Plot} Viable parameter region of axion model considering the DW contribution to axion relic density as estimated by this work (assuming $\Lambda = \Lambda_{\rm QCD}$). The white region indicates that the axion relic abundance is sufficient to account for the observed dark matter as measured by the Planck Observatory ($\Omega_{\rm DM} = (0.12\pm 0.0012)h^{-2}$) \cite{Aghanim:2018eyx}, taking into account both the misalignment mechanism and the DW contribution. The width of the white region presents the uncertainty associated with extrapolation, which expands as $\epsilon$ decreases. Above the black-dashed horizontal line, the DW has not entered the scaling regime before its decay. The yellow area indicates that the produced axion partially contributes to dark matter, while the orange area indicates an overproduction of dark matter. The blue-dashed line represents the prediction of $\Omega_{\rm DM}=\Omega_{a}^{\rm DW} \propto \epsilon^{-1/2}$ from a previous simulation study \cite{Kawasaki:2014sqa}: the area to the lower left of the line indicates overproduction. The result from \cite{Kawasaki:2014sqa} is shown as a thin line as the error bar given there is tiny. The grey/dark grey areas are excluded by BBN constraint and CMB observation, respectively, as DWs must decay prior to the BBN and CMB eras ($t_{\rm decay} < 0.01$s)   \cite{Kawasaki:2004yh,Kawasaki:2004qu}.}
\end{figure}

In this section, we will estimate the contribution of DWs to the relic density of axions based on the results obtained in earlier sections and present the viable parameter space of our model. We will apply our result to the $N_{\rm DW}=2$ ALP model (see Eq.(\ref{Eq: Potential})) with pre-inflationary PQ symmetry breaking (so that cosmic strings are simply absent) as a concrete example. We then present an illustrative analysis that includes the potential contribution of cosmic strings to the axion relic abundance in the Appendix.~\ref{Appendix-QCD}.

The contribution of the standard misalignment mechanism to the axion relic density is found to be negligible compared to the DW contribution in the parameter space of our interest ($\epsilon \lesssim 10^{-3}$): $\rho_{\rm mis}/\rho_{\rm DW} < 1\%$, where $\rho_{\rm mis}\simeq m_a^2 f_a^2/2N_{\rm DW}^2$ is the axion energy density from the misalignment mechanism, and $\rho_{\rm DW}$ is the contribution from DW decay. We thus neglect its contribution in the subsequent discussions. 

The DW contribution to the relic axions is given by the solutions to the evolution equation of motion Eq.(\ref{Eq: EoM rho_2}) and Eq.(\ref{Eq: EoM rho_3}) along with their numerical results in Eq.(\ref{Eq: app rho_2}) and Eq.(\ref{Eq: app rho_3}), respectively. The total axion energy density is $\rho_a = \rho_2 + \rho_3$. In order to estimate the axion relic density $\Omega_a$ from DWs, we numerically fit $\rho_2 + \rho_3$ based on data points in Fig.~\ref{Fig: Velocity}, and extrapolate the result to lower $\epsilon$'s and a wider range of $m_a$. Our fitting result for the DW contribution to $\Omega_a$ is
\begin{align}
\label{Eq: Omega_a DW}
    \notag \Omega_{\rm a}^{\rm DW}h^2 \simeq & \, 0.116 \left(\frac{m_a}{2\times 10^{-4}\,\hbox{eV}}\right)^{-1.50^{+0.02}_{-0.02}} \\ & \times \left( \frac{\Lambda}{400\,\hbox{MeV}} \right)^4 \left(\frac{\epsilon}{ 10^{-4}}\right)^{-1.87^{-0.35}_{+0.44}},
\end{align}
where the uncertainties are fitted within the $m_a$ and $\epsilon$ ranges as given in Fig.~\ref{Relic_NoString_Plot}. The benchmark example with $\Lambda = \Lambda_{\rm QCD}$ is shown in Fig.~\ref{Relic_NoString_Plot}, where the parameter region that predicts the observed axion DM relic density $\Omega_{\rm a} = (0.12\pm 0.0012)h^{-2}$ lies in the white area. We also considered the BBN constraint $t_{\rm decay} < 0.01$s \cite{Kawasaki:2004yh,Kawasaki:2004qu}, and the CMB constraint that DWs should decay before photon decoupling. In addition, the region above the black horizontal dashed line corresponding to $\epsilon = 5\times 10^{-3}$ (also see Fig.~\ref{Fig: FN_epsilon_vary} in Appendix.~\ref{Appendix} indicates that the DW network does not have sufficient time to transition into the scaling regime before its decay. Furthermore, we have fixed $\Lambda = \Lambda_{\rm QCD}$ in Fig.~\ref{Relic_NoString_Plot} as a QCD example, but Eq.~\ref{Eq: Omega_a DW} can apply to general ALPs by varying $\Lambda$, and the constraints shown in Fig.~\ref{Relic_NoString_Plot} related to axion relic abundance would be relieved for smaller $\Lambda$'s.

Fig.~\ref{Relic_NoString_Plot} also includes a comparison between the results from our study and those from the previous 2D simulation for the metastable DW \cite{Kawasaki:2014sqa,Hiramatsu:2012sc}. We use the dashed blue curve to represent the prediction \footnote{The authors in \cite{Kawasaki:2014sqa} used a different notation compared to our Eq.(\ref{Eq: Potential}). Here we used a conversion: $\Xi \simeq \epsilon \frac{m_a^2}{2 f_a^2 N_{\rm DW}^2} $, where $\Xi$ is the bias parameter used in Eq.(3.1) in \cite{Kawasaki:2014sqa}.} of the DW contribution to the axion relic abundance as presented in \cite{Kawasaki:2014sqa}. Both studies have technical limitations that restrict their simulations to relatively large values of $\epsilon \gtrsim \mathcal{O}(10^{-3})$, and extrapolations are made to smaller $\epsilon$ and different $m_a$ values. 
Our estimate of the axion relic abundance for $\epsilon \sim 10^{-4}$ to $10^{-3}$ roughly agrees with that of \cite{Kawasaki:2014sqa}, but a discrepancy becomes increasingly significant as $\epsilon$ decreases. For example, the DW network produces more axion energy density in our finding compared to \cite{Kawasaki:2014sqa} in the range of smaller bias region of $\epsilon \lesssim 10^{-4}$, while results in less axion energy density for $\epsilon \gtrsim 10^{-4}$. In \cite{Kawasaki:2014sqa} the fitting for axion relic density from DWs is $\Omega \propto \epsilon^{-1/2} m_a^{-3/2}$. 
The discrepancy between their and our result may arise from the differences in the fitting models chosen for DW dynamics, especially the DW decay behavior $A_v$. This $A_v$ controls the energy density of DW and explains its decaying process, and thus consequently influences the axion production. We adopt the fitting model described by Eq.(\ref{Eq: A_v Fitting Model}), whereas \cite{Kawasaki:2014sqa} employs a power-law form $A_v \propto t^{1-p}$ with a pressure calibration parameter $p$. This power-law model was investigated in \cite{Larsson:1996sp,Hiramatsu:2010yn}. They analyze the pressure gap between different vacuums, then conclude that the collapse of DWs occurs when the pressure in the true vacuum overcomes the one in the false vacuum, which takes place at $t_{\rm decay} \sim \sigma_{\rm DW}/\Delta V \propto \frac{\epsilon}{m_a}$, where $\Delta V$ represents the difference in potential between the vacua. However, the fitting model described by Eq.(\ref{Eq: A_v Fitting Model}) and Eq.(\ref{Eq: t_decay}) in our work provides a much better fit to our simulation results. These fitting formulae that we used are inspired by the mean-field approximation method analysis in \cite{Hindmarsh:1996xv} as discussed in Sec.~\ref{Sec: Axion Model}.

\section{Conclusion}
This work presents an updated study on the dynamics and evolution of long-lived, metastable axion DWs, with a DW number of $N_{\rm DW} = 2$ as a benchmark. The study incorporates 3D lattice simulations and a semi-analytical approach based on the VOS model. Our analysis includes analyzing the DW evolution dynamics by monitoring the simulation snapshots, and a detailed examination of the axion kinetic energy spectrum. We infer the mechanisms of axion production sourced by the DWs and the corresponding energy loss mechanisms of the DWs. The contribution to the relic abundance of axions from the DW is then derived by numerical fitting and extrapolation, and is found to be significantly greater than that from the misalignment mechanism for a small bias parameter $\epsilon \lesssim 5 \times 10^{-3}$.

Based on the features in the axion energy spectrum obtained from our simulation (see Sec.~\ref{Sec: Features Identified from Simulation}), we identified two distinct components or kinetic energy regimes of the axions: the shorter wave-length axion clouds with resonance around $k \sim 3.68\, m_a$, with larger impact on the small-scale region in the axion spectrum; and the longer wave-length axion mechanical waves with $k \lesssim m_a$. These two features are sourced by different DW dynamics. The axion clouds primarily arise from the flattening motion of the horizon-scale DWs, which smooths the fluctuations on the DW surface while heating (i.e. enlarging the oscillation amplitude of) the background axion fields. On the other hand, the axion mechanical wave is mostly generated by the collapse of the horizon-sized compact DWs which are formed by self-chopping or contraction processes of the super-horizon sized DWs. 

Based on these identified features and the corresponding sources, we derive equations governing the evolution of the DWs, built upon the existing VOS model (for stable DWs) while extending it to incorporate the decay phase of the DWs. By energy conservation, the evolution equation of the DWs is coupled to that of the free axions. By solving these equations numerically, we determine the present-day relic abundance of axions. Our findings align with some earlier literature in terms of the scaling solution, the DW area $A_v$ in Eq.(\ref{Eq: A_v}) and the self-chopping effect in the VOS model. Meanwhile, notable differences are identified and thoroughly discussed. Particularly, our prediction for $\Omega_a (m_a,~\epsilon,~\Lambda)$ takes a different form compared to the results found in \cite{Kawasaki:2014sqa,Hiramatsu:2012sc}, as shown in Eq.(\ref{Eq: Omega_a DW}) and Fig.~\ref{Relic_NoString_Plot}. This discrepancy, which is likely caused by the mathematical fitting model for DW area evolution $A_v$, has potentially significant implications for axion dark matter physics and related experimental probes. Consequently, we predict a larger $\Omega_a$ from the DW decay process in the range of $\epsilon \lesssim 10^{-4}$ compared to the earlier simulation study \cite{Kawasaki:2014sqa}, and a smaller $\Omega_a$ for larger $\epsilon$.

While we directly simulated a simple axion model using the potential described in Eq.(\ref{Eq: Potential}), we have demonstrated that the results can be applied to certain ALP models and the QCD axion models, with a bias parameter $\epsilon \lesssim 10^{-3}-10^{-4}$ that ensure that the DW thickness can be treated as a constant before DWs decay away. See discussion in Sec.~\ref{Sec: Application of our simulation to other models} for the conditions of general applicability, and Sec.~\ref{Sec: Cosmological implication} and Appendix.~\ref{Appendix-QCD} for numerical examples of the application to stringless ALP/QCD models and QCD axion string-wall networks, respectively. In particular, we considered the benchmark of axion mass in the range of $10^{-6} \leq m_a \leq 1 ~$eV with a fixed DW phase transition scale $\Lambda = \Lambda_{\rm QCD}$ as a benchmark.

Notably, our study improves upon existing literature by including the biased potential in the 3D field simulation without relying on approximations such as the PRS algorithm. To ensure efficient simulation with this more accurate treatment, we focus on the benchmark case of $N_{\rm DW}=2$ and decouple the radial mode, which is a reasonable assumption for the relevant time range of DW formation.  It is worth exploring further by considering $N_{\rm DW}>2$ and simulating the full complex scalar field. The dynamics of DWs identified in this study can provide new insights into the physics of axion-like DWs and other types of DWs, such as those arising from GUT models. The updated results on axion DW dynamics presented here are also expected to implications for astrophysical observables related to axion physics, including gravitational wave signals from axion DWs and the formation of axion minihalos as relic overdense energy regions originated from DWs decay. 

\section*{Acknowledgement}
The authors are supported in part by the US Department of Energy under award number DE-SC0008541. The simulation in this work was performed with the UCR HPCC.

\appendix

\section{\label{Appendix-QCD} The application to QCD axion case with string-wall network}

In order to estimate the axion energy density generated by cosmic strings, we employ a conservative estimation outlined in \cite{Buschmann:2021sdq}. They simulated the QCD axion cosmic string evolution in the scenario with a post-inflationary PQ symmetry breaking and a short-lived DW ($N_{\rm DW} = 1$) that formed at the QCD phase transition. We considered $N_{\rm DW} = 2$ in this study, however, we can still apply the cosmic string contribution to the axion field as the result given in \cite{Buschmann:2021sdq,Buschmann:2019icd,Gorghetto:2018myk} to our study. There are two main reasons for this. First, the contribution from cosmic strings that decayed prior to the QCD phase transition should match the simulations presented in references \cite{Buschmann:2021sdq, Buschmann:2019icd, Gorghetto:2018myk}, as this is sourced before DW formation and thus independent of DW details. Second, shortly after the QCD phase transition, the DW tension becomes dominant within the string-DW network, as long as the condition introduced in Eq.~\ref{Eq: String-Wall tension} is met. Therefore in this regime, the string contribution to the axion abundance would be subleading relative to that from the DW, and the possible variance compared to the $N_{\rm DW}=1$ case would be insignificant.

\begin{figure}[t]
\includegraphics[width=0.5\textwidth]{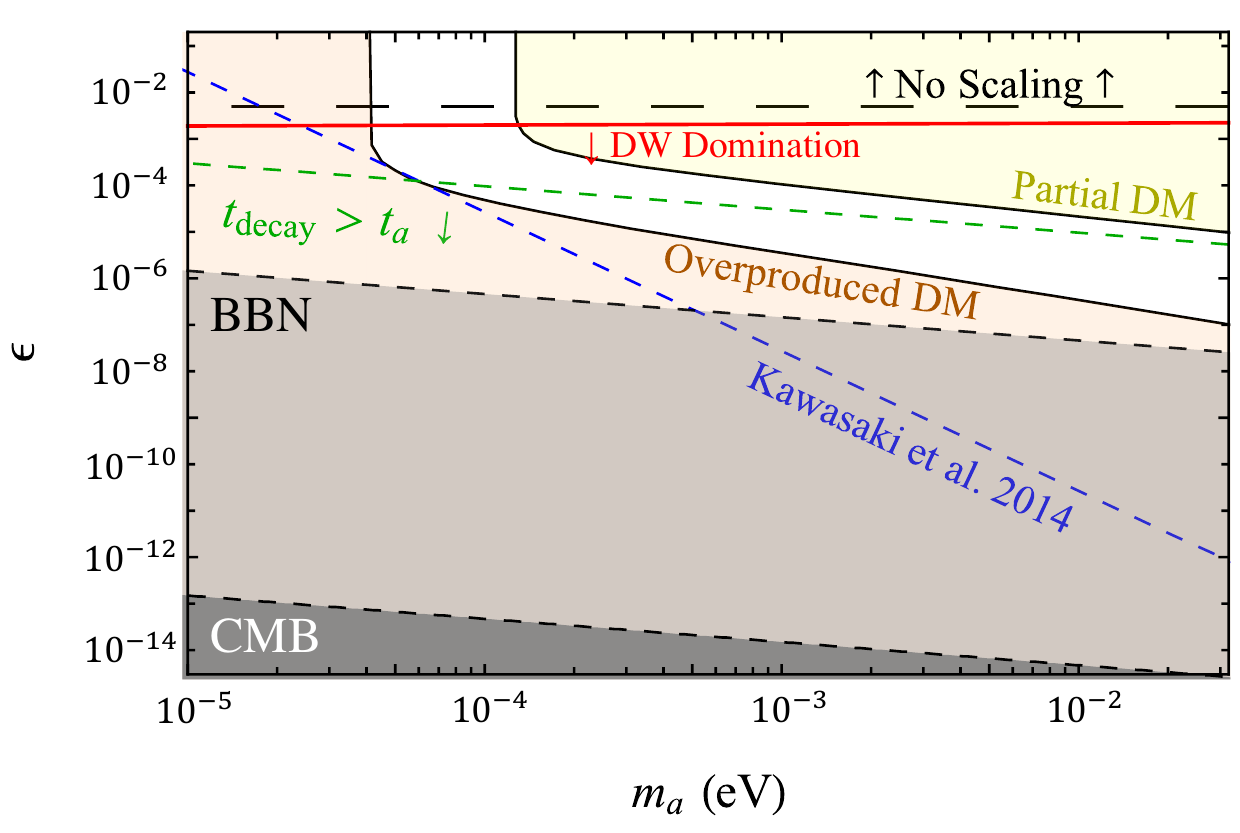} 
\caption{\label{Relic_Plot} Viable parameter region of axion model as estimated by this work, considering both the DW and cosmic strings' contribution to axion relic density ($N_{\rm DW}=2$). The white region indicates that the axion relic abundance is sufficient to account for the observed dark matter as measured by the Planck Observatory ($\Omega_{\rm DW} = (0.12\pm 0.0012)h^{-2}$) \cite{Aghanim:2018eyx}, taking into account both the misalignment mechanism, cosmic string \cite{Buschmann:2021sdq}, and the DW. Above the black-dashed horizontal line, the DW has not entered the scaling regime before its decay. Above the red-solid curve, if a cosmic string exists, string tension dominates the network until the DW decay. Below the green-dashed curve, the thickness of the QCD axion DWs approaches a constant before its collapse, as given by Eq.(\ref{Eq: QCD axion model}). The yellow area indicates that the produced axion partially contributes to dark matter, while the orange area indicates an overproduction of dark matter. The blue-dashed line represents the prediction $\Omega_{a}^{\rm DW} \propto \epsilon^{-1/2}$ from a previous simulation study \cite{Kawasaki:2014sqa}. The grey/dark grey areas are excluded by BBN constraint and CMB observation, respectively, as DWs must decay prior to the BBN and CMB eras ($t_{\rm decay} < 0.01$s) \cite{Kawasaki:2004yh,Kawasaki:2004qu}.}
\end{figure}

As discussed in Sec.~\ref{Sec: Application of our simulation to other models}, our simulation result can be applied to the DW domination period in the QCD axion string-wall network if the following two conditions are met: \\
(1) The domain wall (DW) becomes dominant within the string-wall network ($t> \mu/\sigma_{\rm DW}$, as shown in Eq.(\ref{Eq: String-Wall tension})), and subsequently, it has sufficient time ($\Delta t \sim 10/m_a$) to transition into the scaling regime before its eventual decay. This condition ensures that the influence of the cosmic string on the network becomes negligible and eliminates sensitivity to the initial field distribution, thanks to the attractive (scaling) solution offered by the DW. Furthermore, we specifically consider a scenario with a domain wall number of $N_{\rm DW} = 2$, where two DWs are attached to a single cosmic string. In this case, the DW tension prevails on $t> \mu/\sigma_{\rm DW}$, rendering the impact of the cosmic string negligible. Consequently, the network behaves no differently from a scenario with a single DW ($N_{\rm DW} = 2$ without string) in our simulation. This alignment with the network's evolution after the DW tension dominates is consistent with the findings of our study.\\
(2) The QCD axion domain wall thickness is time-dependent as shown in Eq.(\ref{Eq: QCD axion model}) until cosmic temperature $T = \Lambda_{QCD}$, while our simulation considered a constant thickness. Therefore in order to be self-consistent, the second condition is that the DW network should be long-lived enough to enter the scaling region after $T = \Lambda_{QCD}$.\\
As will be discussed in the following paragraphs, Fig.~\ref{Relic_Plot} shows those two conditions with a red solid line, and a green dashed line, respectively. 

The calculation of the axion energy density produced by cosmic strings in the references \cite{Buschmann:2021sdq, Buschmann:2019icd, Gorghetto:2018myk} considers two distinct contributions from these cosmic structures:\\
(A) Axion radiation during the evolutionary phase, which starts from the cosmic string formation and ends around the QCD phase transition: this contribution arises from the emission of axion radiation by cosmic strings as they evolve. This emission takes place during the earlier phases of cosmic string evolution. This component is particularly significant in determining the axion relic abundance. As per the QCD axion model (referenced as Eq. (\ref{Eq: QCD axion model})), the mass of a single axion particle $m_a(T)$ is inversely proportional to the energy levels at earlier times, following the relationship $m_a(T) \propto T^{-4}$. Consequently, axions with lighter masses are produced during this phase.\\
(B) Decay of the remaining cosmic strings at QCD phase transition: The second contribution stems from the complete decay of the cosmic strings that remain after their evolutionary phase. This decay occurs at the QCD phase transition. \\
\indent The dominant role in determining the axion relic abundance is played by the first contribution (A), where lighter axion particles are produced. This is because the required energy thresholds for axion production are lower during the earlier stages of the universe. The second contribution (B), involving the decay of cosmic strings at the QCD phase transition, accounts for about half of the overall contribution as found in \cite{Buschmann:2021sdq}. 

It is important to note that the specific case being discussed involves a network of strings attached to walls (referred to as a string-walls network). Not all of these strings decay immediately during the QCD phase transition. Some of these strings persist until later stages, and their contribution to the axion abundance is ignored due to the dominance of domain walls (DWs) at that later time, see Eq.(\ref{Eq: String-Wall tension}). 
The remaining strings that have not yet decayed at the QCD phase transition will mostly eventually decay along with the decaying domain walls. Some of them will decay before the DW dominates, but the decay rate should be gradually suppressed by DW tension. The mass of axion particles in this scenario is higher than the mass during the earlier phases, thus fewer axion particles are produced. As a result, the estimation presented in \cite{Buschmann:2021sdq}, which considered the contributions from (A) and from the immediate string decay through (B), could potentially predict a higher axion abundance compared to the string-walls scenario.

As shown in Fig.~\ref{Relic_Plot}, the prediction for the observed axion DM relic density $\Omega_{\rm a} = (0.12\pm 0.0012)h^{-2}$ lies in the white area. The BBN and CMB constraints, scaling region, and a comparison to the early simulation work \cite{Kamionkowski:2014zda} are discussed in Fig.~\ref{Relic_NoString_Plot} and Sec.~\ref{Sec: Cosmological implication}. Furthermore, we present condition (1) as the red line, and condition (2) has been shown as the green dashed line in Fig.~\ref{Relic_Plot}. The prediction of DW-produced axion relic abundance is given in Eq.(\ref{Eq: Omega_a DW}). 

The estimated contribution from cosmic strings is found to be considerably higher than the energy contribution of axions resulting from the misalignment mechanism. Additionally, when domain walls (DWs) have a sufficiently long lifetime, i.e., $\epsilon \lesssim 10^{-3}$, their contribution can surpass that of cosmic strings.

\section{\label{Appendix} Supplemental Data}

\begin{figure*}[t]
 \includegraphics[width=1.03\textwidth]{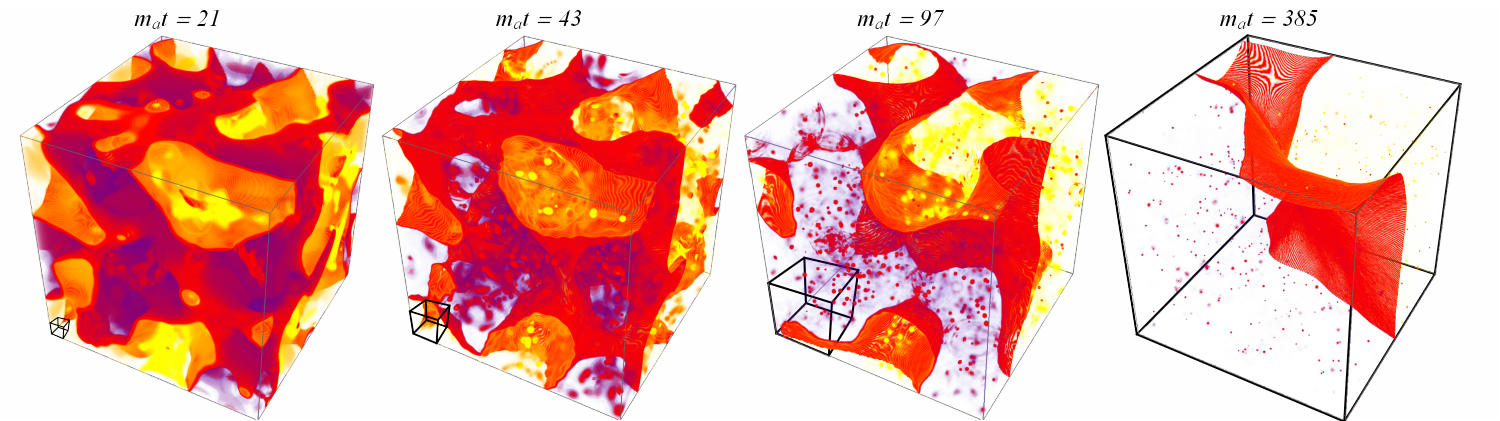}
  \caption{\label{Fig: Simulation no-epsilon} Virtualization of lattice simulation with no biased ($\epsilon=0$) potential from early cosmic time to later (left to right). It is more clear to see the flattening motion of the DW on the right-most and second-right snapshots, in which the DW flats its surface curvature.}
\end{figure*}

\begin{figure}[t]
\includegraphics[width=0.5\textwidth]{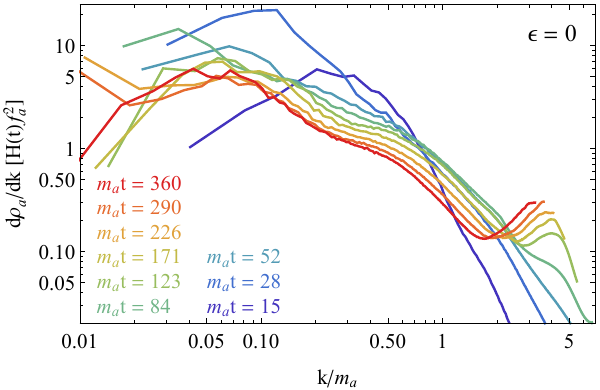} 
\caption{\label{Fig: PRho0} Axion energy density spectrum $\partial \rho_a / \partial k $ versus physical momentum $k$ with no biased potential.}
\end{figure}

\begin{figure}[t]
\includegraphics[width=0.5\textwidth]{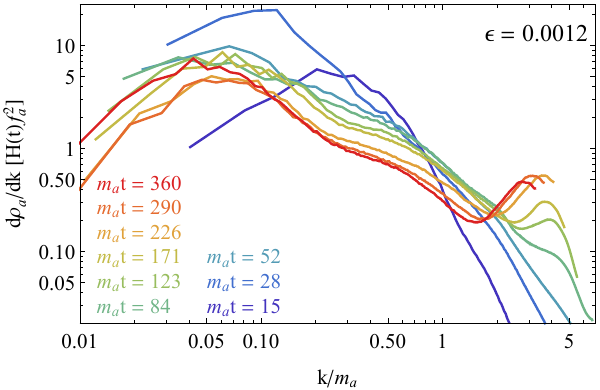} 
\caption{\label{Fig: PRho12} Axion energy density spectrum $\partial \rho_a / \partial k $ versus physical momentum $k$ with bias parameter $\epsilon = 0.0012$.}
\end{figure}

\begin{figure}[t]
\includegraphics[width=0.45\textwidth]{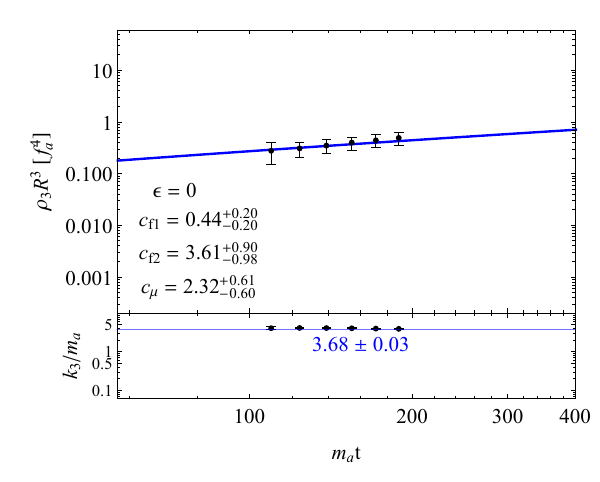} 
\caption{\label{Fig: Fitting epsilon 0} The energy density with $\epsilon = 0$ for the third Gaussian fitting function, see Eq.(\ref{Eq: Fitting_model}). The blue curve presents the prediction of energy loss model Eq.(\ref{Eq: EoM rho_3}). We excluded the data from the early time $m_a t<100$ because its amplitude is too small, and the fitting result has big uncertainty. The later time data $m_a t > 200$ has also been excluded because the peak of $\rho_3$ is out of $k_{\rm N_y}$, and we are thus not able to fit the model nicely.   }
\end{figure}
\begin{figure}[t]
\includegraphics[width=0.45\textwidth]{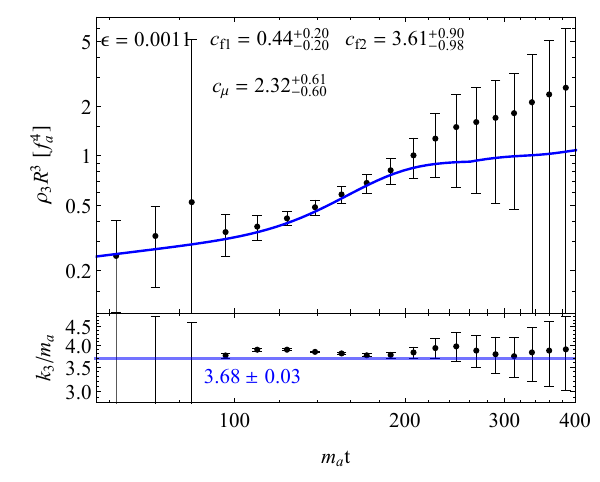} 
\caption{\label{Fig: Fitting epsilon 11} The energy density with $\epsilon = 0.0011$ for the third Gaussian fitting function as given in Eq.(\ref{Eq: Fitting_model}). The blue curve presents the prediction of energy loss model Eq.(\ref{Eq: EoM rho_3}).  }
\end{figure}

\begin{figure}[t]
\includegraphics[width=0.45\textwidth]{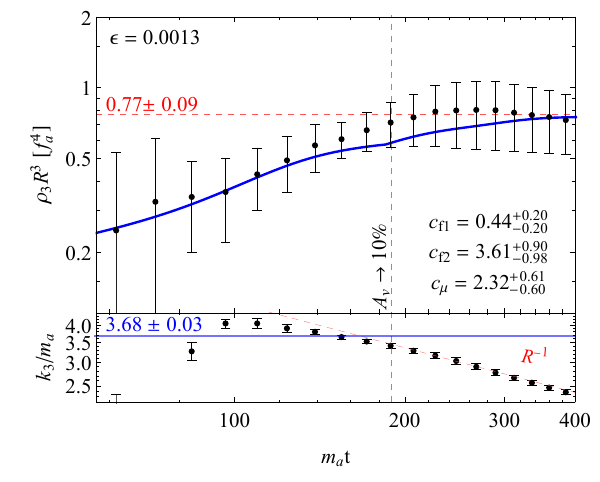} 
\caption{\label{Fig: Fitting epsilon 13} The energy density with $\epsilon = 0.0013$ for the third Gaussian fitting function as given in Eq.(\ref{Eq: Fitting_model}). The blue curve presents the prediction of energy loss model Eq.(\ref{Eq: EoM rho_3}).  }
\end{figure}

\begin{figure}[t]
\includegraphics[width=0.45\textwidth]{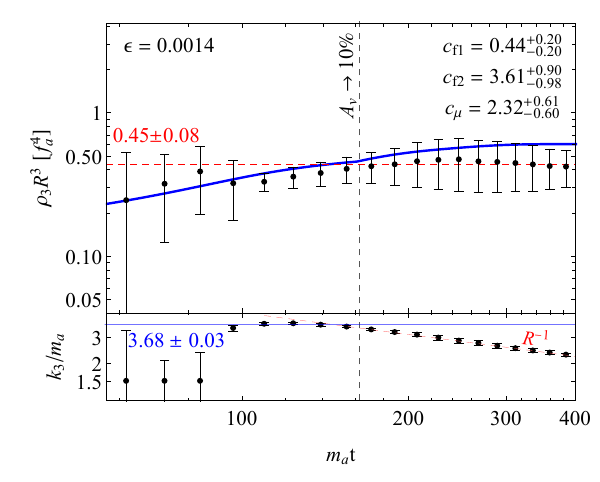} 
\caption{\label{Fig: Fitting epsilon 14} The energy density with $\epsilon = 0.0014$ for the third Gaussian fitting function as given in Eq.(\ref{Eq: Fitting_model}). The blue curve presents the prediction of energy loss model Eq.(\ref{Eq: EoM rho_3}).  }
\end{figure}

\begin{figure}[t]
 \includegraphics[width=0.499\textwidth]{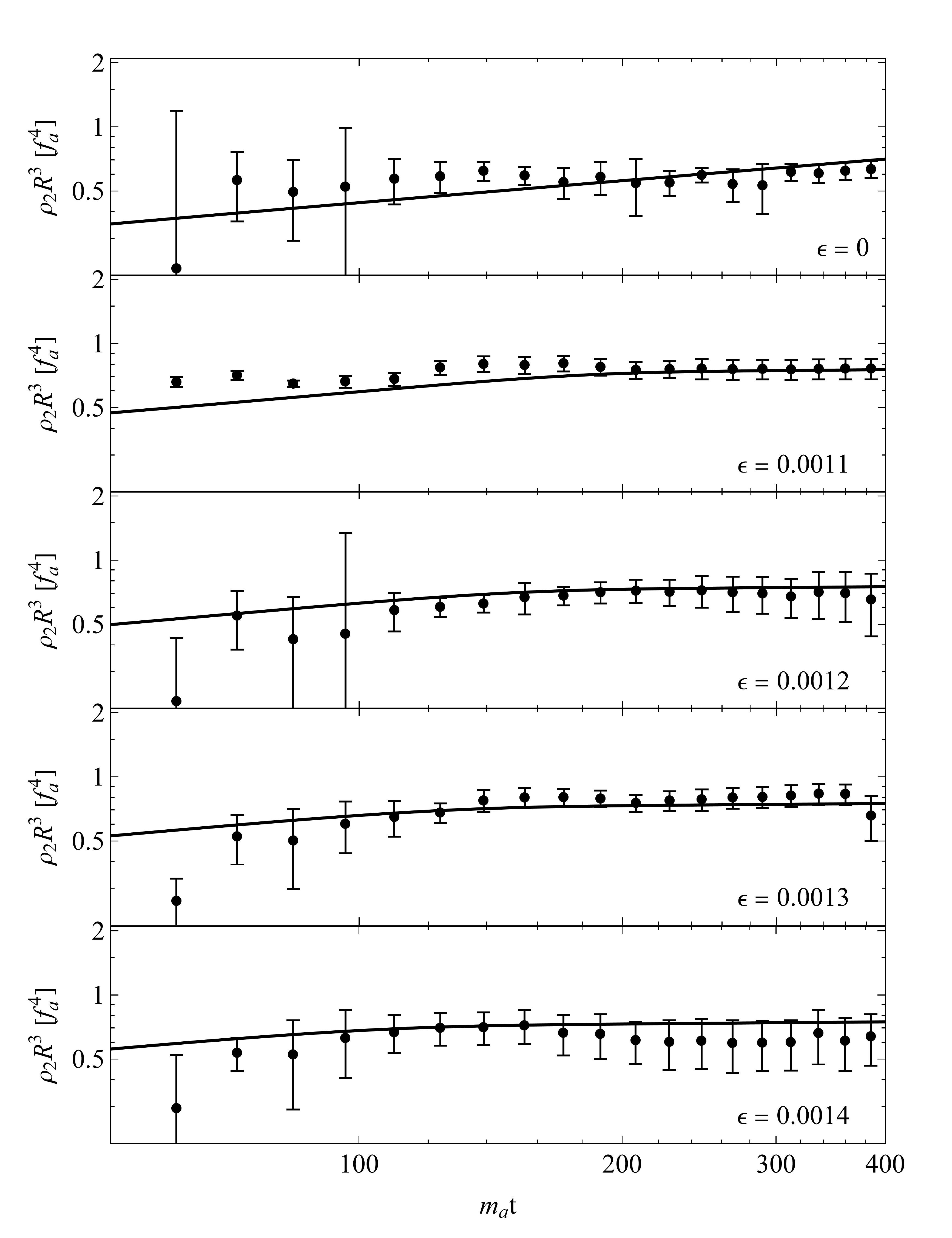}
  \caption{\label{Fig: Rho_Peak_2} The energy density of the second Gaussian fitting function as given in Eq.(\ref{Eq: Fitting_model}) where we provide a variation of $\epsilon$ as marked in the figure. The black curve presents the prediction of energy loss model Eq.(\ref{Eq: EoM rho_2}). }
\end{figure}

\begin{figure}[t]
 \includegraphics[width=0.49\textwidth]{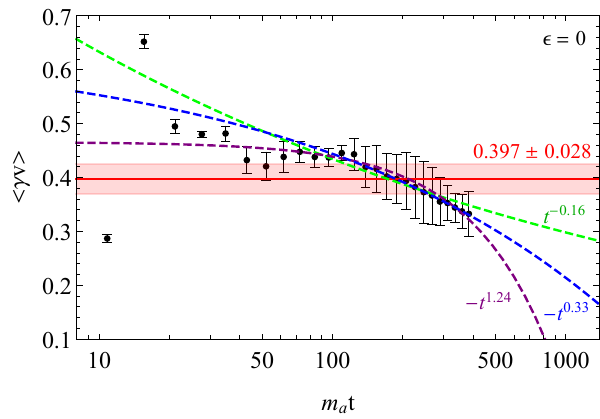}
  \caption{\label{Fig: Velocity_peak_0} Averaged DW velocity with a relativistic factor $<\gamma v>$ versus $m_at$ with benchmark $\epsilon = 0$. The black error bars are observation data in the simulation. The red area presents a constant fit. The dashed purple curve fits with whole time ranges. The dashed green curve fits with $m_a t \geq 15$ which corresponds to the scaling regime. And the dashed blue curve fits with $m_a t \geq 20$ that excludes the high-velocity data point at $m_a t \sim 15$ where the network just right entered the scaling regime (see Fig.~\ref{Fig: AVt_mat}, the $A_v$ becomes a constant).  }
\end{figure}

\begin{figure}[t]
\includegraphics[width=0.5\textwidth]{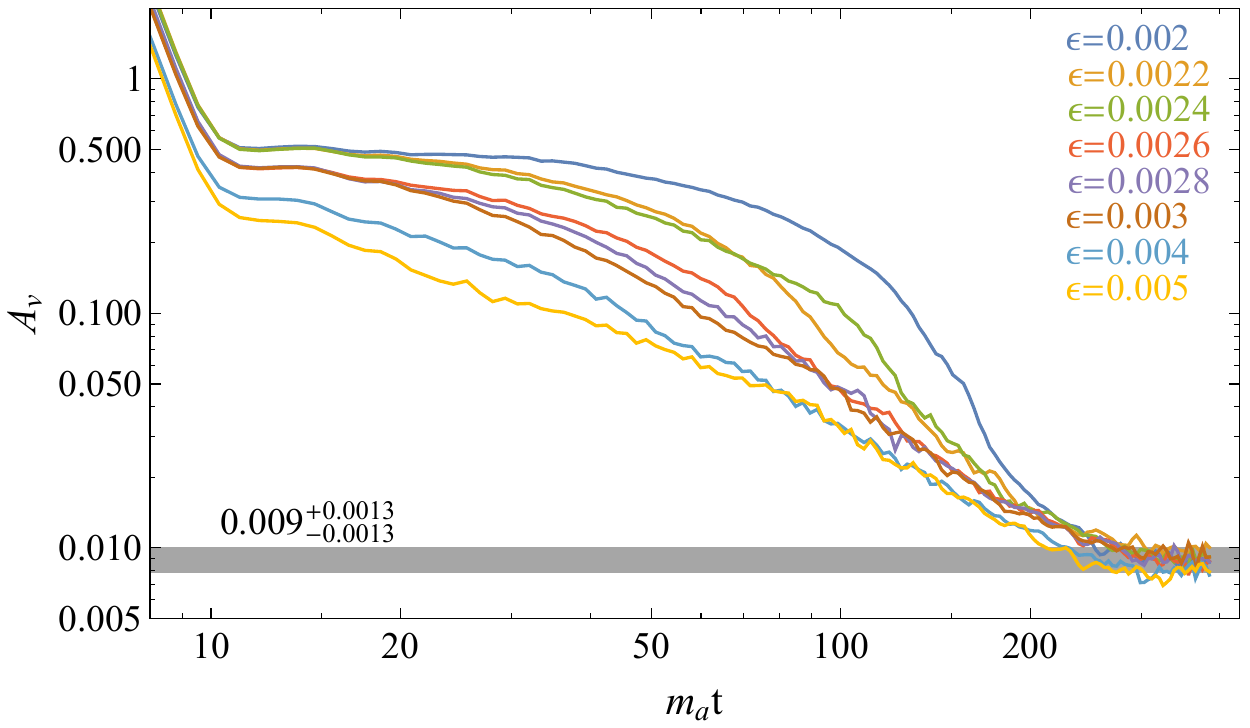} 
\caption{\label{Fig: FN_epsilon_vary} Domain wall area parameter to simulation cosmic time with varying bias parameter $\epsilon$. All the benchmarks converge to $A_v = 0.009^{+0.0012}_{-0.0012}$. As the yellow curve, $\epsilon = 0.005$, the DW enters the scaling regime with a short period $ 11 \lesssim m_a t \lesssim 17$, then decays shortly after. We thus conclude that the DW has enough time to enter the scaling regime for $\epsilon \lesssim 0.005$. }
\end{figure}

In this appendix, we provide supplementary data for the following purposes:\\
$\bullet$ We present a simulation animation for a no biased potential $\epsilon = 0$ in Fig.~\ref{Fig: Simulation no-epsilon}. The right-most and the second-right snapshots clearly present the flattening motion of the DW. \\
$\bullet$ Axion kinetic energy density spectrum with benchmarks $\epsilon = 0$ and $\epsilon = 0.0012$ are given in Fig.~\ref{Fig: PRho0} and Fig.~\ref{Fig: PRho12}, respectively.\\
$\bullet$ The model fits for $\rho_3$ with benchmarks $\epsilon=0$, $\epsilon = 0.0011$, $\epsilon = 0.0013$, and $\epsilon = 0.0014$  are shown in Fig.~\ref{Fig: Fitting epsilon 0}, Fig.~\ref{Fig: Fitting epsilon 11}, Fig.~\ref{Fig: Fitting epsilon 13}, and Fig.~\ref{Fig: Fitting epsilon 14}, respectively. \\
$\bullet$ The model fits for $\rho_2$ with different benchmarks are shown in Fig.~\ref{Fig: Rho_Peak_2}.\\
$\bullet$ Fig.\ref{Fig: Velocity_peak_0} displays the various potential model fit options for the DW velocity $\langle \gamma v \rangle$ when $\epsilon = 0$, which corresponds to fitting the first term in Eq.(\ref{Eq: gamma velocity}). The interpolation results for later times $m_a t \gg 1$ are significantly influenced by different assumptions made about the data, such as when the network enters the scaling regime. In this particular study, we assumed that the network enters the scaling regime when $A_v$ becomes constant, i.e.$m_a t$, as shown in Fig.~\ref{Fig: AVt_mat}.\\
$\bullet$ We increase the bias parameter $\epsilon$ from $0.002$ to $0.005$ to verify a limitation of $\epsilon$ that whether the DW network enters into the scaling region before its decay in our simulation. Fig.~\ref{Fig: FN_epsilon_vary}.

\newpage
\bibliographystyle{apsrev4-1}
\bibliography{References}

\end{document}